\documentclass[fleqn,usenatbib,useAMS]{mnras}
\usepackage{pdflscape}	
\usepackage[T1]{fontenc}
\usepackage{ae,aecompl}
\usepackage{breakurl}
\usepackage{natbib}
\newcommand{\ic}{\textit{I\/}$_\mathrm{C}$}
\newcommand{\rc}{\textit{R\/}$_\mathrm{C}$}
\newcommand{\jb}{\textit{J\/}}
\newcommand{\h}{\textit{H\/}}
\newcommand{\ks}{\textit{K\/}$_\mathrm{s}$}
\newcommand{\jhks}{\textit{JHK\/}$_\mathrm{s}$}
\newcommand{\gaia}{\textit{Gaia\/}}
\newcommand{\iras}{\textit{IRAS\/}}
\newcommand{\spitzer}{\textit{Spitzer\/}}
\newcommand{\akari}{\textit{Akari\/}}
\newcommand{\iso}{\textit{ISO\/}}
\newcommand{\iraf}{{\sevensize IRAF}}
\newcommand{\wise}{\textit{WISE\/}}
\newcommand{\herschel}{\textit{Herschel\/}}
\newcommand{\msun}{\,M$_{\sun}$}
\newcommand{\msunyr}{\,M$_{\sun}$\,yr$^{-1}$}
\newcommand{\lsun}{\,L$_{\sun}$}
\newcommand{\av}{\textit{A\/}$_\mathrm{V}$}

\newcommand{\lbol}{\textit{L\/}$_\mathrm{bol}$}
\newcommand*{\swap}[2]{#2#1}
\usepackage{amsmath}	
\usepackage{amssymb}	
\usepackage{graphicx}
\usepackage{url}
\begin{document}

\title[2MASS~22352345+7517076]{The outbursting protostar 2MASS~22352345+7517076 
and its environment}

\author[M. Kun et al.]{M. Kun$^{1}$\thanks{E-mail: kun@konkoly.hu}, P.
\'Abrah\'am$^{1}$, J. A. Acosta Pulido$^{2}$, A. Mo\'or$^{1}$, and T.
Prusti$^{3}$ \\
$^{1}$Konkoly Observatory, Research Centre for Astronomy and Earth Sciences, 
Hungarian Academy of Sciences,\\ 
H-1121 Budapest, Konkoly Thege \'ut 15--17, Hungary \\ 
$^{2}$Instituto de Astrof\'{\i}sica de Canarias, Avenida V\'{\i}a L\'actea,
38205 La Laguna, Tenerife, Spain \\
$^{3}$ Directorate of Science, ESTEC--ESA, PO Box 299, 
2200 AG, Noordwijk, The Netherlands}
\date{Accepted XXX. Received YYY; in original form ZZZ}
\pubyear{2017}
\label{firstpage}
\maketitle

\begin{abstract} 
We studied the Class~I protostar 2MASS~22352345+7517076 whose dramatic brightening between the \textit{IRAS\/}, \akari\ and \wise\ surveys was reported by \citet{onozato2015}. 2MASS~22352345+7517076 is a member of a small group of low-mass young stellar objects, associated with IRAS~22343+7501 in the molecular cloud Lynds~1251. The \iras, \iso, \spitzer, \akari, \herschel, and \wise\ missions observed different stages of its outburst. Supplemented these data with archival and our own near-infrared observations, and considering the contributions of neighbouring sources to the mid-infrared fluxes we studied the nature and environment of the outbursting object, and its photometric variations from 1983 to 2017. The low-state bolometric luminosity \lbol $\approx 32$\lsun\ is indicative of a 1.6--1.8\msun, 1--2$\times10^5$ years old protostar. Its 2-\micron\  brightness started rising between 1993 and 1998, reached a peak in 2009--2011, and started declining in 2015.  Changes in the spectral energy distribution suggest that the outburst was preceded by a decade-long, slow brightening in the  near-infrared. The actual accretion burst occurred between 2004 and 2007. We fitted the spectral energy distribution in the bright phases with simple accretion disc models. The modelling suggested an increase of the disc accretion rate from $\sim 3.5\times10^{-7}$\msunyr\ to $\sim 1.1\times10^{-4}$\msunyr. The central star accreted nearly $10^{-3}$\msun, about a Jupiter mass during the ten years of the outburst. We observed H$_2$ emission lines in the \textit{K\/}-band spectrum during the fading phase in 2017. The associated optical nebulosity RNO~144 and the Herbig--Haro object HH~149 have not exhibited significant variation in shape and brightness during the outburst. 
\end{abstract}


\section{Introduction}

Star formation begins in the densest regions of interstellar molecular clouds. 
The earliest phases of mass accumulation take place in the most opaque parts of
the cloud cores. Stellar mass is build up from accretion discs, resulting
from the rotation of the collapsing cloud core. Accretion discs act as mass 
reservoirs. Mass accretion from the disc onto the star is driven by disc 
instabilities, therefore is a strongly variable process. 
Theoretical considerations \citep[e.g.][]{Vorobyov2015} suggest that the masses of Sun-like stars build up during a sequence of accretion bursts. The most powerful bursts occur in the embedded phase of star formation, when the disc is fed by the gas from the collapsing envelope.  Outbursts of the optically visible young stars can be observed in the FU~Orionis and EX~Lupi type stars. Mechanisms of mass accumulation and disc heating, leading to accretion bursts, are discussed in several theoretical papers  \citep[e.~g.][]{Bell1994,Bell1995,Vorobyov2006,Zhu2009,Bae2014}. Each scenario postulates the presence of an envelope, therefore FUors may represent a transition from embedded to optically visible young stars. According to the scenario developed by \citet{,Zhu2009}, episodic outbursts can be explained by accumulation of matter in the inner disc region due to angular momentum rearrangements in the outer disc, resulting from gravitational instabilities. The increasing degree of ionization at the high-temperature inner disc region activate other (magneto-rotational, thermal) instabilities which forward gas from the disc onto the star. 

Outbursting protostars \citep[Class~0 and Class~I young stellar objects, YSOs,][]{greene94} are younger siblings of FUors, observable only in the infrared. Duration of outbursts and their effects on the central star and its circumstellar environment, reflected in the near-infrared spectra \citep{Connelley2018}, distinguish these objects from FUors and EXors. Infrared observations of the recent decade \citep[e.g.][]{contreras1} revealed dozens of embedded protostars with various amplitudes and time scales of outburst. Several of them exhibited near-infrared spectra similar to optically visible FU~Ori and EX~Lupi type eruptive stars \citep{contreras2}. \citet{Kuffmeier2018} performed a detailed simulation of protostellar evolution in various environments and found that infall on to the circumstellar disc may trigger gravitational instabilities in the disc at distances  around 10 to 50~AU, leading to strong accretion bursts which typically last for about 10 to 100 yr, consistent with typical orbital times at the location of the instability. Outbursting protostars are excellent laboratories for constraining the physical processes of episodic accretion. They are, however, rare and difficult to find. Only a few of them were examined in detail \citep[e.g.~OO~Ser, V2775~Ori, and HOPS~383,][respectively]{kospal2007,fischer2012,safron2015}. 

The target of the present paper, 2MASS~22352345+7517076, belongs to the class of the embedded eruptive YSOs. It is associated with IRAS~22343+7501, a protostellar source embedded in the molecular cloud Lynds~1251. \citet{RD95} identified a cluster of five near-infrared sources associated with IRAS~22343+7501 (RD95~A, B, C, D, and E). The cluster is a source of several molecular outflows \citep{sato1989,nikolic2003,kim2015}, the Herbig--Haro jet HH~149 \citep{balazs1992}, and radio continuum jet sources \citep[VLA~6, 7, and 10,][]{reipurth2004}. At optical wavelengths it is associated with the faint, red reflection nebula RNO~144 \citep{cohen1980}. 

The dramatic brightening of \textit{IRAS\/}~22343+7501 at mid-infrared wavelengths between the \textit{IRAS\/} (1983), \akari\ (2006), and \wise\ surveys was reported by \citet{onozato2015}. They established that the outbursting star was RD95\,D, coinciding with 2MASS~22352345+7517076 and with VLA\,6, and found that most of the brightening occurred between 2006 and 2010. Their near-infrared observations have shown that the \ks\ magnitude of the outbursting star was some 4~mag brighter in 2013 than the 2MASS \ks, measured in 1999. The mid-infrared \iras, \akari, and \wise\ fluxes, however, contain fluxes of other members  of the \iras~22343+7501 group, therefore \citet{onozato2015} have not attempted to untangle the true nature of the outbursting star and the outburst phenomenon.  

In order to characterize the central star and the outburst of 2MASS~22352345+7517076 in more detail we analysed archival data from \iras\ to \textit{NEOWISE--R\/}, and performed near-infrared spectroscopic and photometric observations in 2016 and 2017. We corrected all mid-infrared fluxes for the contribution of neighbouring sources. We describe the new and archival observational data in Section~\ref{Sect_data}. Light curves, colour--magnitude and colour--colour diagrams, as well as spectral energy distributions are presented in Sect.~\ref{Sect_res}, and discussed in Sect.~\ref{Sect_disc}. We summarize our results in Sect.~\ref{Sect_sum}.

\section{Data}
\label{Sect_data}

 \subsection{Distance of Lynds~1251}
 \label{Sect_dist}
 
Literature values of the distance of L1251 are $300\pm50$\,pc \citep{KP93} and  $330\pm30$\,pc  \citep{balazs2004}. Parallaxes of optically visible members of Lynds~1251, published in \gaia\ Data Release~2 \citep{gaia2018}, allow us to derive an improved distance of this star-forming region. The average parallax of the 15 known members of L1251, included in \gaia~DR2, and the standard deviation of the average result in a distance of $350^{+46}_{-38}$\,pc. We adopt the new distance during the present work.

\subsection{Photometric and spectroscopic observations}
\label{Sect_obs}

We list in this Section the data available for the outbursting protostar, including our own and archival data chronologically. Original observations and data reductions are described in detail. 

\paragraph*{\textit{IRAS\/}, 1983.} The 12, 25, 60, and 100\,$\mu$m fluxes of IRAS~22343+7501 are composite of the mid- and far-infrared fluxes of several sources, revealed by near-infrared observations within the field of view of the \textit{IRAS\/} detectors \citep{RD95}. We estimated and subtracted the contribution of neighbouring sources (see Sect.~\ref{Sect_fluxcor}) from the \textit{F\/}$_{12}$ and \textit{F\/}$_{25}$ fluxes, listed in the \iras\ Faint Source Catalog, and use the corrected fluxes during this work.

\begin{figure*}
\centerline{\includegraphics[width=5cm]{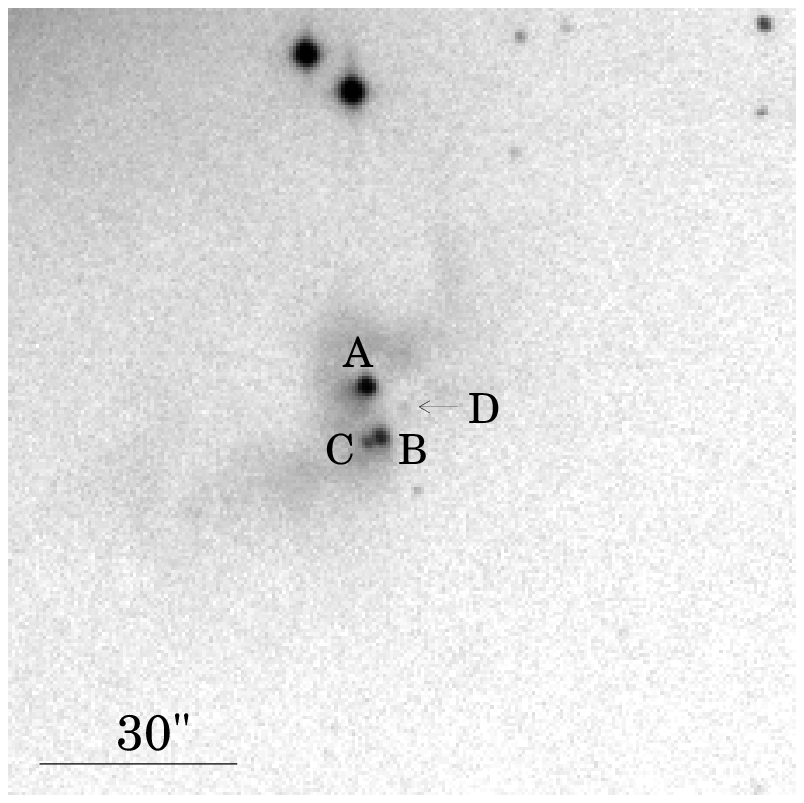}
\includegraphics[width=5cm]{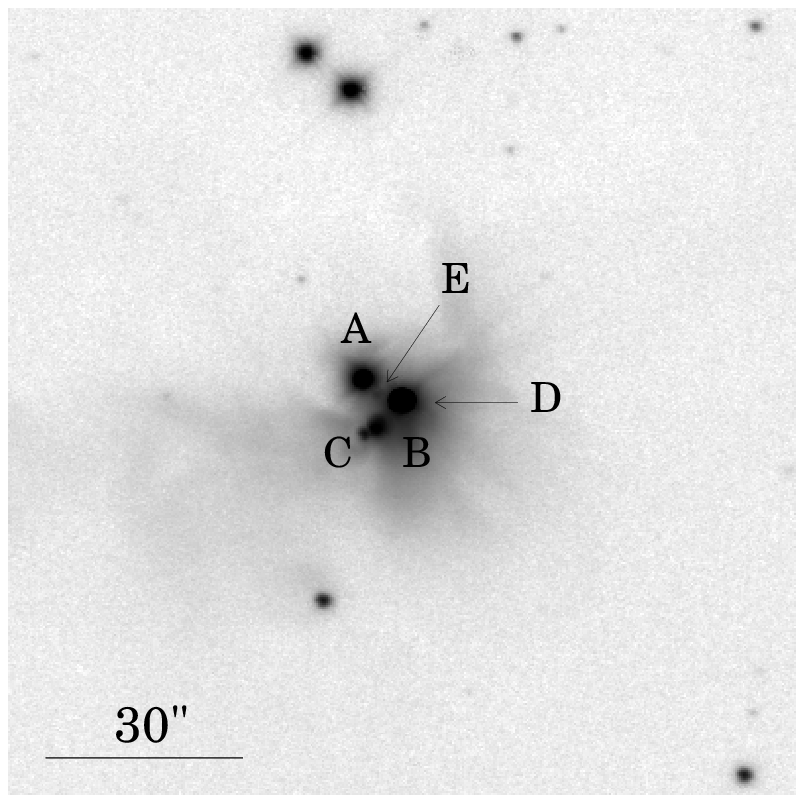}
\includegraphics[width=5cm]{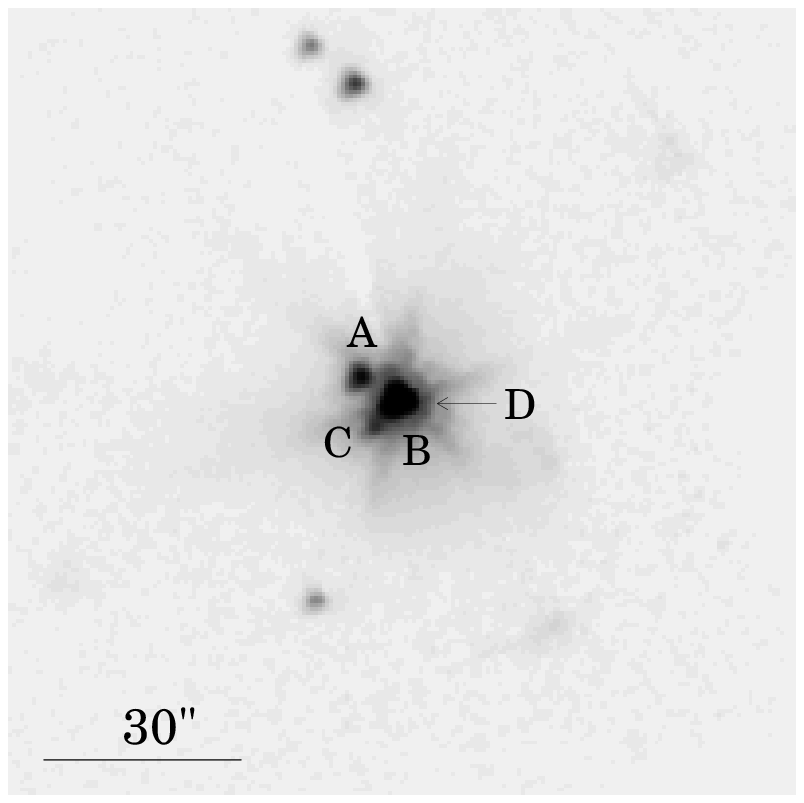}}
\centerline{\includegraphics[width=5cm]{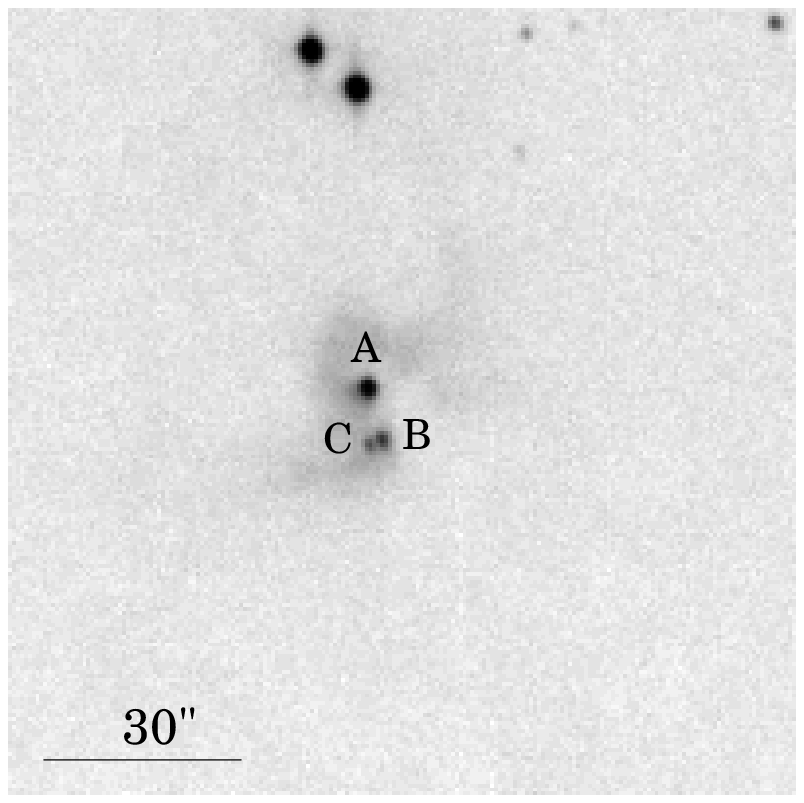}
\includegraphics[width=5cm]{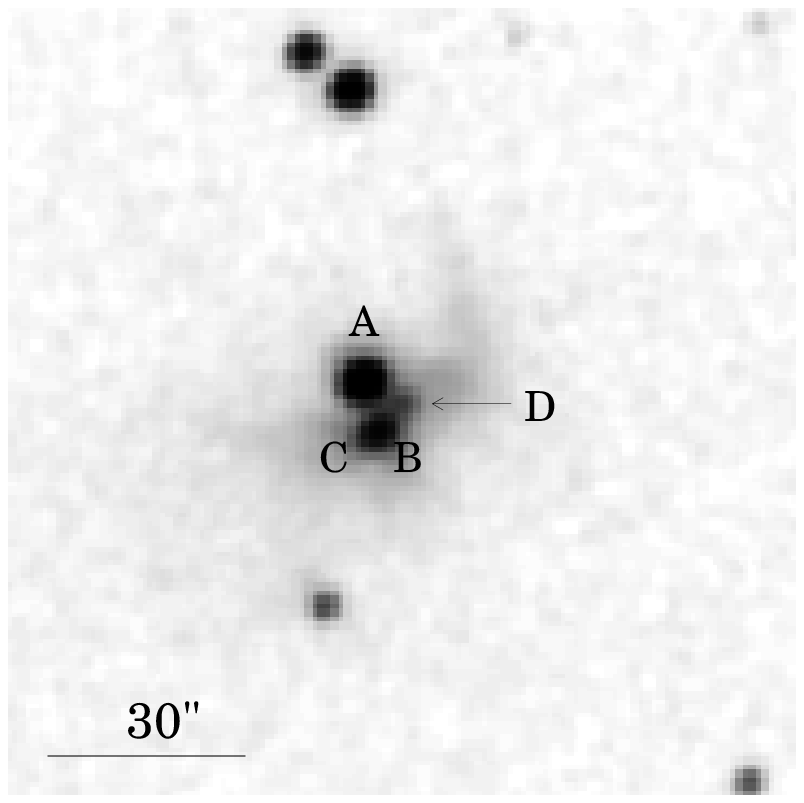}
\includegraphics[width=5cm]{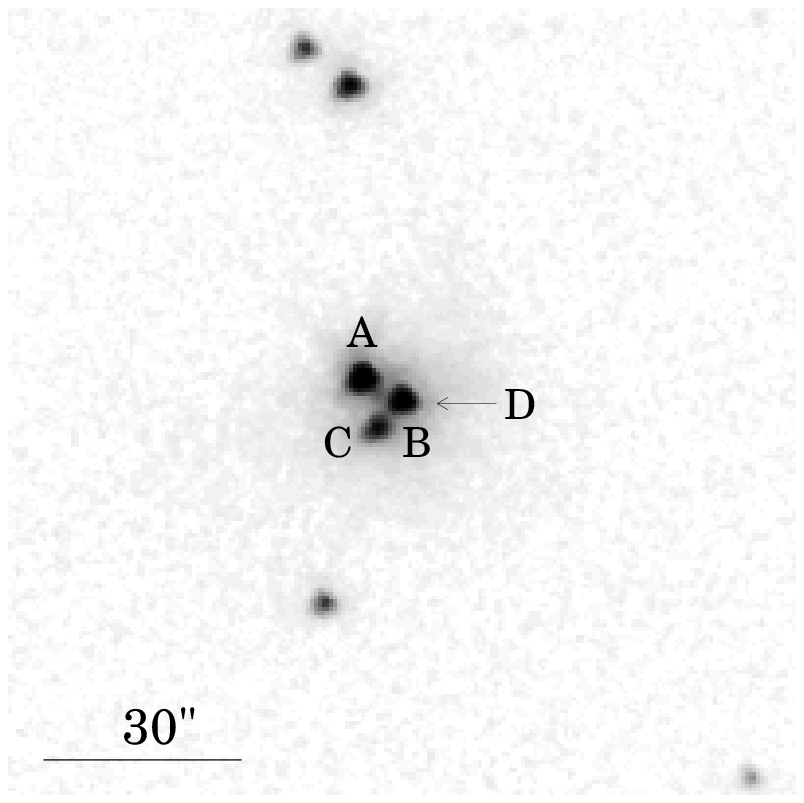}}
\caption{Lower row: Appearance of IRAS~22343+7501 before the outburst. \textit{Left\/}: optical $I_\mathrm{C}$-band image obtained in 1999. The stars RD95\,A, RD95\,B,  RD95\,C are labelled as A, B, and C. The nebula RNO\,144 can be seen around these central stars. \textit{Middle\/}: 2MASS $K_\mathrm{s}$-band image (1999). Our target is the faint star D; \textit{Right\/}: \spitzer\ IRAC 3.6-\micron\ image obtained in 2004. Upper row: The same field after the outburst. \textit{Left\/}:  \ic-band image obtained in 2009; \textit{Middle\/}: $K_\mathrm{s}$-band image observed with the CFHT WIRCam in 2011; \textit{Right\/}: IRAC 3.6-\micron\ image observed in 2009 September. The outbursting star is positioned at the centre of each $2\arcmin\times2\arcmin$ image.} 
\label{Fig_images}
\end{figure*}

\paragraph*{Pre-outburst optical observations, 1990 and 1999.} Optical images of the region were taken in 1990 November 18 with the 3.5-m telescope of the Calar Alto Observatory through narrow-band H$\alpha$ and \ion{S}{ii}, and broad-band red
filters. The observations resulted in the discovery of the optical jet HH~149  
\citep{balazs1992}. We obtained optical images of the same region through Johnson \textit{R\/} and \textit{I\/} filters using the CAFOS instrument, installed on the 2.2-m telescope of the Calar Alto Observatory on 1999 August 7. The images were reduced in \iraf. We examine them to reveal possible changes in the shape and brightness of the optical nebula during the outburst. The lower left panel of Fig.~\ref{Fig_images} shows a $2\arcmin\times2\arcmin$ part of the  \textit{I\/}-band image.

\paragraph*{Pre-outburst near-infrared and submillimeter data, 1993.} \citet{RD95} published near-infrared \jb\ and \textit{K\/}$^\prime$-band magnitudes, measured on images observed on 1993 August~30 with the Redeye detector, installed on the Canada--France--Hawaii Telescope (CFHT) for five sources, denoted as A, B, C, D, and E, associated with IRAS~22343+7501. The outbursting source is RD95\,D, the second faintest of the five objects, clustered within the field of view of the \iras. They estimated a visual extinction of $32\pm5$~mag toward the line of sight of the sources. These authors also observed IRAS~22343+7501 at 0.85, 1.1, and 1.3 millimeter, using the UKT14 detector on the James Clerk Maxwell telescope.

\paragraph*{\iso\ observations, 1996.} \iras~22343+7501 was observed at 4.5 and 12.0\,$\mu$m with the ISOCAM instrument \citep{cesarsky1996} on board the \textit{Infrared Space Observatory\/} (\iso) on 1996 December 31. The data were reduced with the CAM Interactive Analysis Software V5.0 \citep[CIA][]{Ott1997}. Pipeline-processed data are available in the \iso\ Archive\footnote{\url{www.iso.esac.esa.int/ida/index.html}}. The sources identified in a $6\times6$ mosaic of images, centred on RA(2000)=$22^\mathrm{h}35^\mathrm{m}23\fs45$ and Dec(J2000)=+75\degr17\arcmin06\arcsec\ are listed in Table~\ref{Table_isocam}. ISOC\,223522.5+751705 corresponds to our target. It was saturated in the 12~\micron\ image. 

 \begin{table*}
 \caption{ISOCAM 4.5-\texorpdfstring{\micron}{mu} sources in the region of \textit{IRAS\/}~22343+7501, observed on 1995 December 31. }
 \label{Table_isocam}
 \begin{center}
 \begin{tabular}{lccl}
 \hline
 \noalign{\smallskip}
 Id. &  $F_{4.5}\pm$e($F_{4.5}$) & $F_{12}\pm$e($F_{12})$ & Cross Id. \\
      & (mJy) & (mJy) & \\
\noalign{\smallskip}
\hline
\noalign{\smallskip}
ISOC\,223411.1+751810  & 821.1$\pm$24.6 & 932.8$\pm$18.7 & [KP93] 2-1 \\	
ISOC\,223501.7+751758  & 24.6$\pm$3.4  & 26.6$\pm$3.7 & [KLC2015]\,8 \\ 
ISOC\,223503.3+751820  & 17.3$\pm$3.3 &  25.3$\pm$3.7 & [KLC2015]\,9 \\ 
ISOC\,223515.8+751848  & 61.3$\pm$ 5.8 & 104.8$\pm$8.4 &  [KP93] 2-39,[KLC2015]\,12 \\
ISOC\,223522.7+751707  & 260.9$\pm$19.6 & 1864.1$\pm$28.0$^*$ & [KLC2015]\,1, RD95\,D  \\
ISOC\,223523.9+751712  & 728.6$\pm$21.9 & $\cdots$ &[KLC2015]\,3, RD95\,A \\
ISOC\,223524.5+751757  & 111.3$\pm$7.2 & 85.1$\pm$5.5 & [KLC2015]\,13 \\	 
ISOC\,223526.1+751638  &  44.7$\pm$3.4 & 48.1$\pm$7.5 & [KLC2015]\,14 \\
ISOC\,223526.0+751802  &  28.2$\pm$3.0 & 27.9$\pm$4.1& RD95\,5 \\
ISOC\,223605.0+751832  & $\cdots$ &  361.4$\pm$10.8 & [KLC2015]\,17 \\
 \hline 
\end{tabular}
\end{center}
\smallskip
\flushleft{\small $^*$Affected by saturation.}
\end{table*}

L1251 was also observed with the ISOPHOT detector of \iso\ on 1996 December~31 at 100, 120, and 200\,\micron, using the C100 camera at 100 and 120\,\micron\ and the C200 camera at 200\,\micron. Maps of the cloud were obtained with the P22 astronomical observing template mode \citep{Laureijs2003}. The data reduction was performed using the ISOPHOT Interactive Analysis Software Package V10.0 \citep[PIA][]{Gabriel1997}. We followed in detail the processing scheme described in \citet{Burgo2003}. The source's flux density was determined by fitting the profile with the footprint of a point source on top of an extended baseline \citep{abraham2000}. 

\paragraph*{2MASS 1999.} The outbursting star was observed on 1999 October~11.  2MASS~J22352345+7517076 was detected only in the \ks\ band with photometric quality flag `E', and lower magnitude limits are given for the \jb\ and \textit{H\/} bands. The 2MASS \textit{K}$_\mathrm{s}$ image of the \iras~22343+7501 region is shown in the lower middle panel of Fig.~\ref{Fig_images}. Our target is the faint source D at the centre. Sources A and B correspond to 2MASS~22352497+7517113 and 2MASS~22352442+7517037, respectively.

\paragraph*{SHARC-II 2003} \iras~22343+7501 was observed at 350\,\micron\ in 2003 September \citep{suresh2016} with the SHARC-II instrument, having angular resolution of 10\arcsec. The size of object is $27.1\arcsec\times25.3\arcsec$, corresponding to $9500\times8870$~AU at 350\,pc. 

\paragraph*{\spitzer\ IRAC 2004.} The molecular clump L1251\,C, containing IRAS~22343+7501, was observed by the IRAC camera of the  \textit{Spitzer Space Telescope\/} \citep{fazio2004}, as part of \textit{The Cores to Discs\/} (c2d) Legacy programme \citep{Evans2003}. IRAC observations were performed on 2004 October 18. \citet{kim2015} identified 19 YSOs in the clump, and presented improved flux values for 3.6, 4.5, 5.8, and 8.0\,\micron. The sources IRS\,1, IRS\,2, and IRS\,3 correspond to the near-infrared sources RD95\,D, RD95\,B, and RD95\,A, respectively. The \spitzer\ data have shown IRS\,1 to be a Class~I YSO, whereas IRS\,2 and IRS\,3 are Class~II sources \citep{greene94}. The 3.6-\micron\ \textit{c2d\/} image of the region is shown in the lower right panel of Fig.~\ref{Fig_images}. 

\paragraph*{\spitzer\ MIPS 2004.} Our target was also observed with the Multiband Imaging Photometer for Spitzer \citep[MIPS;][]{Rieke2004} at 24 and 70\,{\micron} on 2004 September 24, as part of the \textit{c2d\/} programme. We downloaded the MIPS data from the Spitzer Heritage
Archive\footnote{\url{http://sha.ipac.caltech.edu/applications/Spitzer/SHA/}}.
At 24\,{\micron} we used the pipeline (S18.12.0) produced post-BCD mosaic
in our analysis. The data processing of the MIPS 70\,\micron\ measurement
was started with the basic calibrated data (BCD) images. As a first step,
we removed residual artifacts from the images by applying column spatial
filtering and time-median-filtering as described by \citet{Gordon2007}.
The improved BCD data then were co-added and corrected for array
distortions using the MOsaicking and Point source Extraction
tool \citep[MOPEX,][]{Makovoz2005}. During the latter step permanently damaged 
pixels and data flagged in the BCD mask files were also discarded.
IRAS~22343+7501 is strongly saturated in the 24-{\micron} image.
Assuming that our target is a point source at this wavelength
we used the \texttt{repair\_saturated} routine of the \texttt{starfinder}
tool \citep{Diolaiti2000} to replace the core of the star's image with a
template representing an estimate of the model point-spread function (PSF).
The PSF was constructed following \citet{Engelbracht2007}.
Finally we performed PSF photometry to extract the flux density of the source.
At 70\,{\micron} we applied aperture photometry. The aperture radius was set to 
35{\arcsec}, while the background was estimated in a sky annulus from 39 to 65{\arcsec}. The appropriate aperture correction factor, valid for sources with a
temperature of 60\,K, was taken from \citet{Gordon2007}.
By analyzing observations of bright stars and asteroids with the MIPS 70-{\micron}
array \citet{paladini2009} found that photometry of sources
brighter than 2\,Jy could be severely affected by non-linearity at high count
rates. To correct this effect we used the formula proposed by \citet{paladini2009} for aperture photometry in their figure\,6 (straight line model). Uncertainties were computed by adding quadratically the internal error and the absolute calibration uncertainty (4 and 7\% for 24 and 70\,{\micron} data, respectively; MIPS Data Handbook).

\paragraph*{SMA 1.3-mm 2007} Three members of \iras~22343+7501 were resolved at 1.3\,mm by the Submillimeter Array \citep{kim2015}. The dust condensation associated with 2MASS~22352345+7517076 had a deconvolved size of $3.3\arcsec\times3.1\arcsec$ (corresponding to $1155\times1083$\,AU at 350\,pc).  

\paragraph*{\akari\ IRC and FIS, 2006--2007.} \iras~22343+7501 was detected by both the InfraRed Camera \citep[IRC,][]{onaka2007} and Far Infrared Surveyor \citep[FIS,][]{kawada2007} instruments on board the \akari\ Infrared Space Satellite \citep{akari}, operated between 2006 April and 2007 August. Fluxes at 9\,\micron\ and 18\,\micron\ are found in the \akari\ IRC Point Source Catalog. Far-infrared fluxes at 65, 91, 140, and 160\,{\micron} were adopted from Version~2.0 of the \akari\ FIS Bright Source Catalog, available in the \akari\ science archives \citep{yamamura2016}. The field of view of both instruments extended to more than one member of the group of YSOs identified by \citet{RD95}. We corrected the 9 and 18-\micron\ IRC fluxes for the contribution of RD95\,A and RD95\,B (see Sect.~\ref{Sect_fluxcor}). 

\paragraph*{\akari\ IRC Post-Helium near-infrared observations, 2008.}
\iras~22343+7501 was observed with the IRC on 2008 August 12, during the Phase~3 (post-Helium) Mission of \akari\ (proposal id.: AFSAS, target id: 1640320, PI: M. Ueno). Pipeline-processed flux-calibrated images, observed through N2 (2.4\,\micron), N3 (3.2\,\micron), and  N4 (4.1\,\micron) filters, as well as a slitless grism spectrum over the 2.5--5.0\,\micron\ region are available in the \akari\ archives \citep{yamashita2016}. The field of view was about 9.1$\times$10.0, 9.3$\times$10.0 and 9.5$\times$10.0\,arcmin for the N2, N3, and N4 filters, respectively, and the pixel size was 1.46\arcsec. Two sets of images were obtained, a short ($\sim$ 4.67\,s), and a long-exposure one (44.4\,s). Our target was strongly saturated in the long-exposure images. We performed aperture photometry on the pipeline-processed short-exposure images. We measured the fluxes of stars within the field of view in 2-pixel apertures, and the sky background on 2-pixel wide annuli around the apertures. Then we applied aperture correction, determined by measuring several isolated field stars in a series of apertures from 2 to 12 pixel radii. 
Results of the photometry are listed in Table~\ref{Tab_nirc}. 

\begin{table}
\caption{Near-infrared fluxes of 2MASS~22352345+7517076 measured in the \akari\ IRC images, obtained on 2008 August 12.}
\label{Tab_nirc}
\begin{center}
\begin{tabular}{lccc}
\hline
\noalign{\smallskip}
Band & Wavelength & Flux$\pm$eFlux \\
     & (\micron) & (Jy) \\
     \hline
\noalign{\smallskip}
N2 & 2.4 & 0.388$\pm$0.060 \\
N3 & 3.2 & 2.511$\pm$0.226 \\
N4 & 4.1 & 5.846$\pm$0.188 \\
\hline
\end{tabular}
\end{center}
\end{table}

\begin{figure}
\centering \includegraphics[width=\columnwidth]{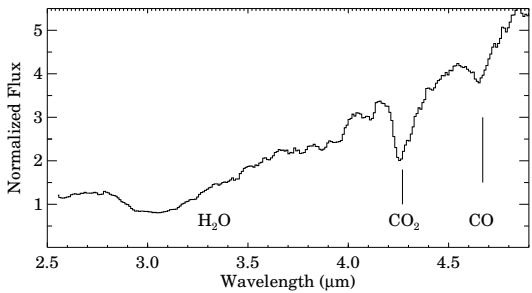}
\caption{The IRC spectrum of IRAS~22343+7501, observed on 2008 August 12.} 
\label{Fig_ircsp}
\end{figure}

The slitless spectrum of IRAS~22343+7501 was obtained through the {\em NG} grism, having a dispersion of 0.0097\,\micron\,pix$^{-1}$. Our target is saturated in the spectroscopic image. To get a qualitative insight into the spectral appearance of the object we extracted one-dimensional spectrum using the outer, unsaturated columns of the pipeline-reduced, wavelength-calibrated, background-subtracted image. The spectrum is shown in Fig.~\ref{Fig_ircsp}. 

\paragraph*{\spitzer\ IRAC Post-Helium observations, 2009-2010.} The T~Tauri star [KP93]\,2-2 \citep{KP93}, located at an angular distance of some 3\,arcmin from IRAS~22343+7501, was a target of monitoring observations using the \textit{Spitzer Space Telescope\/} in the post-helium phase (PID: 60167, PI: P.~\'Abrah\'am). The region was observed with a daily cadence between 2009 Sep~18--Sep~23, and between 2010 Jan~9--Jan~16. The IRAC instrument at 3.6 and 4.5\,\micron\ was used in full-array mode with exposure times of 0.4\,s per frame. 2MASS~22352345+7517076 is saturated in each image. We determined the flux of the star on the post-BCD mosaic images, using annular apertures that avoided the central, saturated part of the stellar image, using the `phot' task of \iraf. The centre of the saturated image was adjusted by measuring its position compared to the nearby sources RD95\,A and RD95\,B. Absolute fluxes of unsaturated stars were determined by measuring their fluxes in a 4-pixel aperture, the sky background on an annulus between the 20th and 33th pixel, and applying aperture correction, given in the \textit{IRAC Instrument Handbook\/}\footnote{\url{http://irsa.ipac.caltech.edu/data/SPITZER/docs/irac/iracinstrumenthandbook/27/}}. Then we measured their fluxes in apertures of 3, 4, 5, and 6 pixel radii, and obtained the fluxes in annular apertures by subtracting the smaller aperture fluxes from the larger ones. The ratio of the flux measured in the annulus between \textit{r1\/} and \textit{r2\/} to the total flux, $F_{r1-r2}/F_{total}$, was derived for the unsaturated stars in the mosaic, and the average of this ratio was used to convert the flux of the target star, measured in the annular aperture, to total flux. Variations of the point spread function within the image are accounted for in the uncertainty of the $F_{r1-r2}/F_{total}$ ratio. The uncertainties were estimated as quadratic sum of the photometric and flux ratio error. Finally we averaged the results, obtained from the annuli between 4--5 and 5--6 pixel radii, and applied colour correction according to the \textit{IRAC Instrument Handbook\/}. The results are listed in Table~\ref{Tab_irac}. We show a 3.6-\micron\ image, averaged from all images obtained in 2009 September in the upper right panel of Fig.~\ref{Fig_images}. 

\begin{table}
\caption{\spitzer\ IRAC fluxes of 2MASS~22352345+7517076 at 3.6 and 4.5\,\micron.}
\label{Tab_irac}
\begin{center}
\begin{tabular}{lccc}
\hline
\noalign{\smallskip}
Date of Obs. & MJD & F$_{3.6}\pm eF_{3.6}$ &  F$_{4.5}\pm eF_{4.5}$  \\ 
yyyymmdd &   & (Jy) & (Jy) \\
\noalign{\smallskip} \hline
 \noalign{\smallskip}
20041018 &  53296.00 &   ~1.550$\pm$0.320$^*$ &  ~2.510$\pm$0.390$^*$ \\ 
20090918 &  55092.16 &   9.380$\pm$0.684 & 17.461$\pm$0.164  \\ 
20090919 &  55093.11 &  11.283$\pm$0.188 & $\cdots$        \\   
20090920 &  55094.07 &  10.488$\pm$0.128 & 18.138$\pm$0.642  \\ 
20090921 &  55095.96 &   9.051$\pm$0.290 & 19.380$\pm$0.278  \\ 
20090921 &  55095.17 &   9.770$\pm$0.526 & 20.007$\pm$0.880  \\
20090923 &  55097.29 &   8.938$\pm$0.157 & 17.617$\pm$0.102  \\ 
20090923 &  55097.99 &  10.040$\pm$0.109 & 17.847$\pm$0.560  \\ 
20100110 &  55206.58 &  12.107$\pm$0.245 & 18.242$\pm$0.250  \\ 
20100111 &  55207.49 &  11.524$\pm$0.182 & 16.939$\pm$0.430  \\ 
20100112 &  55208.18 &  10.608$\pm$0.108 & 17.452$\pm$0.217  \\ 
20100113 &  55209.32 &  11.707$\pm$0.356 & 18.977$\pm$0.560  \\ 
20100114 &  55210.66 &  10.370$\pm$0.318 & 15.658$\pm$0.902  \\ 
20100115 &  55211.92 &  14.723$\pm$0.102 & 17.496$\pm$0.433  \\ 
20100116 &  55212.99 &   9.012$\pm$0.359 & 16.080$\pm$0.402 \\ 
\hline
\end{tabular}
\end{center}
\flushleft{\small
Results from \textit{c2d} measurements, \citet{kim2015}.}
\end{table}

\paragraph*{Post-outburst optical observations, 2009.} We obtained optical images of the region through Cousins \rc\ and \ic\ filters using the CAFOS instrument, installed on the 2.2-m telescope of the Calar Alto Observatory on 2009 October 10 and 14. Three images per filter and per night were taken with total exposure times of 180\,s for the \ic\ and 360\,s for the \rc\ band. The images were bias- and flatfield-corrected and coadded in \iraf. The \ic\ image, shown in the upper left panel of Fig.~\ref{Fig_images}, resulted from coadding all \ic\ images obtained during both nights. A very faint star is visible at the position of 2MASS~22352345+7517076 in this image, whereas the object was invisible in the coadded \rc\ image.

\paragraph*{\herschel\ PACS far-infrared photometric data, 2009--2011.} L1251 was observed with the PACS instrument of the \herschel\ Space Observatory, as part of  The Gould Belt Survey (PI: P. Andr\'e) between 2009 December 28 and 2011 June 29 at 70, 100, and 160\,\micron. The point source fluxes were published separately for the three bands in the PACS Point Source Catalogs \citep{Marton2017}.

\paragraph*{\wise\ data 2010--2017.} 2MASS~22352345+7517076 was observed by the \textit{Wide-field Infrared Survey Explorer\/} \citep{Wright2010} at
3.4, 4.6, 12.0, and 22.0\,$\mu$m (\textit{W1\/}, \textit{W2\/}, \textit{W3\/}, and \textit{W4\/} bands, respectively) on 2010 Feb 4--6 and in the \textit{W1\/}\ and \textit{W2\/} bands on 2010 Aug 13--16. We downloaded all time-resolved observations from the \textit{AllWISE\/} Multiepoch Photometry Table \citep{wise2013}. The source is affected by saturation in the \textit{W1\/} and \textit{W2\/} bands.
Its brightness is, however, below the limit  where the number of the unsaturated pixels are sufficient for reliable profile-fit photometry (2.0 and 1.5~mag for the \textit{W1\/} and \textit{W2\/} bands, respectively, see Sect. 6.3 of the \wise\ Explanatory Supplement). We computed the average of all high-quality (ph$\_$qual equal A or B) profile-fit magnitudes and converted the average magnitudes into fluxes. Colour corrections were applied following the \wise\ Explanatory Supplement\footnote{\url{http://wise2.ipac.caltech.edu/docs/release/allsky/expsup/sec4_4h.html}}. We note that the \textit{AllWISE\/} Source Catalog, based on all available measurements, gives a low-quality $[W2]$ magnitude for our target. 

Since the beam sizes of \wise\ are approximately 6\,arcsec in both the \textit{W1\/} and \textit{W2\/} bands, contamination from the neighbouring point sources RD95\,A and RD95\,B had to be taken into account. We used the \spitzer\ IRAC fluxes of these pre-main-sequence stars, listed by \citet{kim2015} to estimate their contribution to the \textit{W1\/} and \textit{W2\/} fluxes. We subtracted the sum of the \spitzer\ 3.6 and 4.5-\micron\ fluxes of RD95\,A and RD95\,B from the \textit{F}$_{W1}$ and \textit{F}$_{W2}$ fluxes of the \wise\ source, respectively. In the errors, we added in quadrature 10\,\% to account for the uncertainty of the contributing neighbouring sources.  The contributions of these neigbouring T~Tauri stars to the fluxes in the \textit{W3\/} and \textit{W4\/} bands were estimated from their spectral slopes (see Sect.~\ref{Sect_fluxcor}).

Further observations in the \textit{W1\/}\ and \textit{W2\/} bands were obtained between 2014 February and 2017 August during the \textit{NEOWISE\/} reactivation mission \citep{Mainzer2014}. The profile-fit magnitudes listed in the \textit{NEOWISE-R\/} Single Exposure Source Tables indicate brightening of the source between 2010 and 2014. According to the \textit{NEOWISE\/} Explanatory Supplement\footnote{\url{http://wise2.ipac.caltech.edu/docs/release/neowise/expsup/}}, however, profile-fit brightnesses of saturated sources in this database are systematically overestimated, therefore these data have to be regarded unreliable, in spite of their good formal photometric quality indicators.

\paragraph*{Archival CFHT data from 2011 and 2014.}
The region of L1251 containing our target was observed on 2011 Sep 19 and 
2014 Jul 11 in the \ks\ band with the WIRCam near-infrared camera 
installed on the Canada--France--Hawaii Telescope (PI: K.-W. Hodapp). We downloaded the images from the CFHT Science Archive and performed photometry  using \iraf. 2MASS~22352345+7517076 was saturated in each image. We determined its \ks-band magnitude using the non-saturated outer part of the stellar image. For calibration we measured in the same way several stars in the field of view which have high-quality 2MASS data. The results are shown in Table~\ref{Tab_nir}. The \ks\ image observed in 2011 is displayed in the upper middle panel of Fig.~\ref{Fig_images}. 

\paragraph*{Submillimeter data 2012--2014.} L1251 was included in  the JCMT Gould Belt Legacy Survey, and observed with the SCUBA-2 instrument between 2012 March 30 and 2014 October 24. The angular resolution of SCUBA-2 observations was  9.6\arcsec\ and 14.1\arcsec\ at 450\,$\mu$m and 850\,$\mu$m, respectively. Derived fluxes, size, temperature and mass of the source corresponding to IRAS~22343+7501 (source 61) were published by \citet{Pattle2017}. 
 
\begin{figure}
\centering \includegraphics[width=\columnwidth]{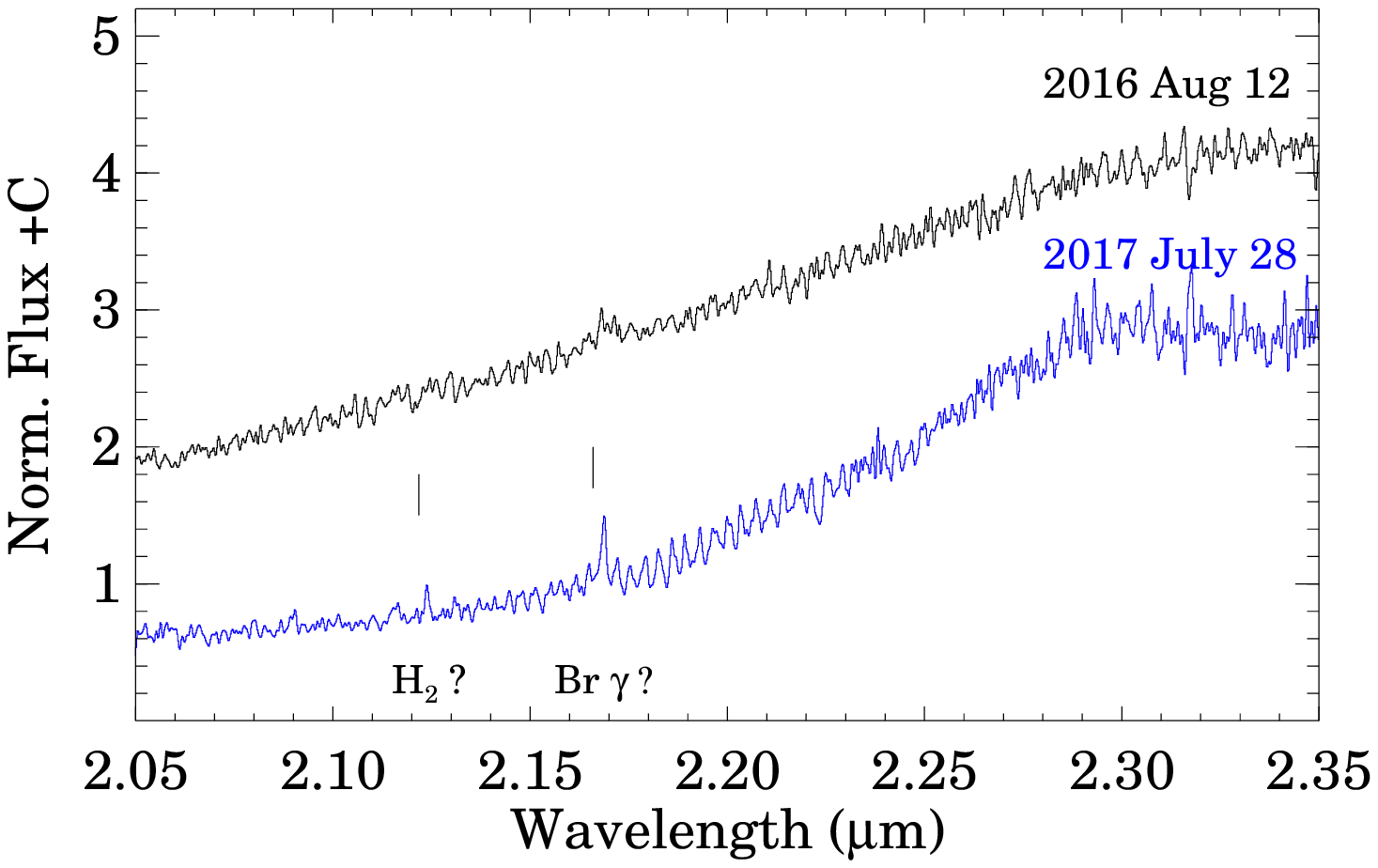} 
\centering \includegraphics[width=\columnwidth]{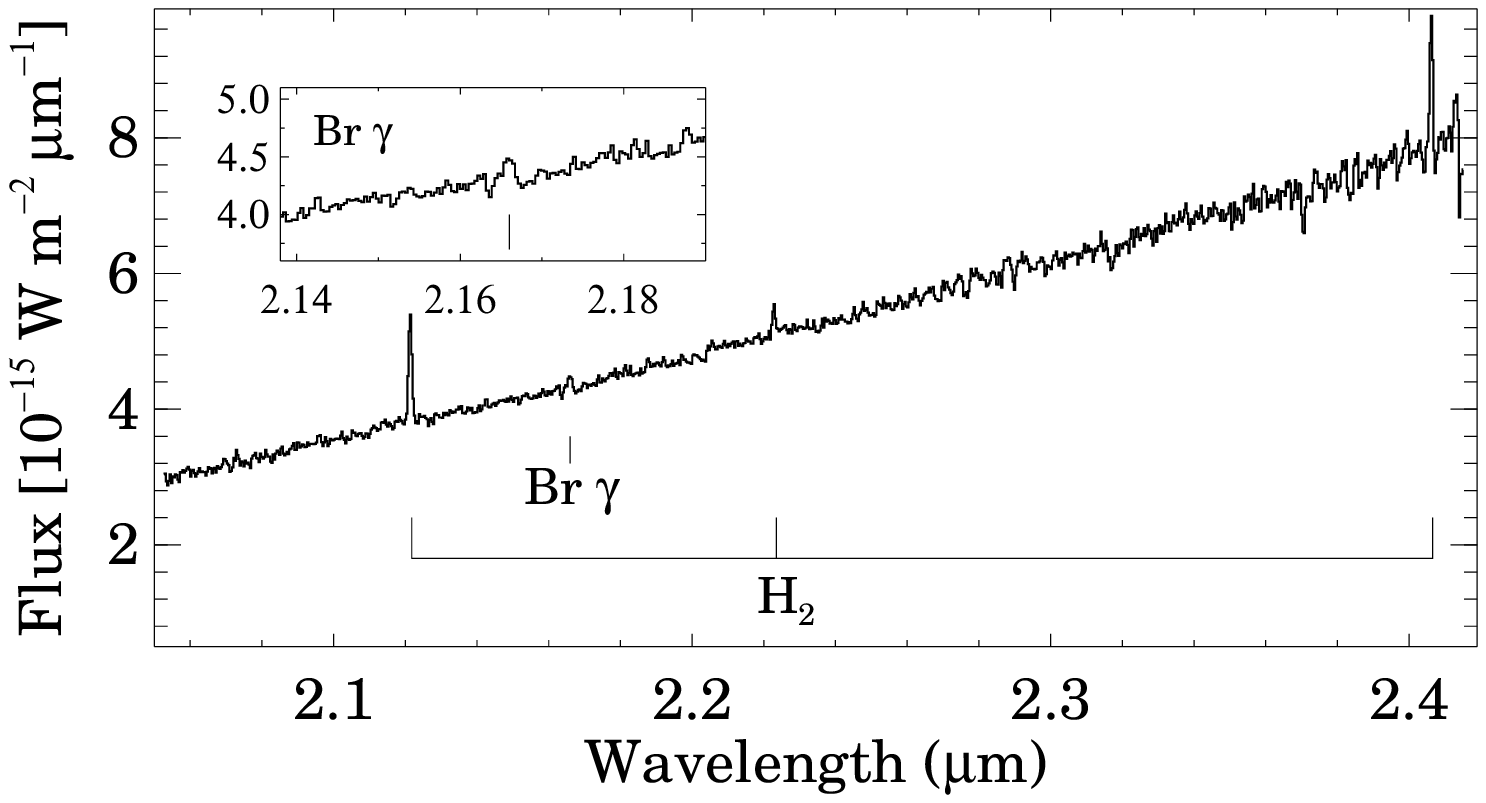}
\caption{Upper panel: \textit{K\/}-band spectra of 2MASS~22352345+7517076 obtained with NOTCam on 2016 August 12 (black) and 2017 July 28 (blue). Lower panel: \textit{K\/}-band spectrum obtained with LIRIS on 2017 October 10.} 
\label{Fig_kspec}
\end{figure}

\paragraph*{NOTCam photometric and spectroscopic observations in 2016 and 2017.} We obtained \jhks\ images and \textit{K\/}-band spectra of
2MASS~22352345+7517076 on 2016 Aug 12 and 2017 Jul 28 using the Wide-Field Camera of the NOTCam instrument installed on the Nordic Optical Telescope in the Observatorio del Roque de los Muchachos in La Palma, Spain. The spectral resolution of the instrument was $R \approx 2500$, using a 0.6\,\arcsec\ slit. The spectra were obtained  with AB-BA dithering and ramp-sampling readout mode in 2016, and the AB3 dither pattern was used in 2017. The total exposure time was 240\,s in 2016 and 540\,s in 2017. Spectra of Xenon and Argon lamps were observed for wavelength calibration, and that of a halogen lamp for flatfielding. The O9.5 type star XZ~Cep, located at an angular distance of some eight degrees from the target was observed for telluric corrections. The spectra were reduced and analysed in \iraf. The \textit{K\/}-band spectra are shown in the upper panel of Fig.~\ref{Fig_kspec}.

For the imaging observations nine-point dithering and ramp-sampling readout 
mode were applied for the \jb\ and \h\ bands, and 9-point dither, reset-read-read readout mode for the \ks-band. The total exposure times were 540\,s in the \jb, 36\,s in the \h, and 10\,s in the \ks\ band in 2016, and the same figures were 1080\,s, 90\,s, and 27\,s in 2017. We reduced the images and applied aperture photometry in \iraf. The instrumental magnitudes were calibrated using 2MASS magnitudes of several stars in the field of view. The results are listed in Table~\ref{Tab_nir}.

\begin{figure}
\centering \includegraphics[width=\columnwidth]{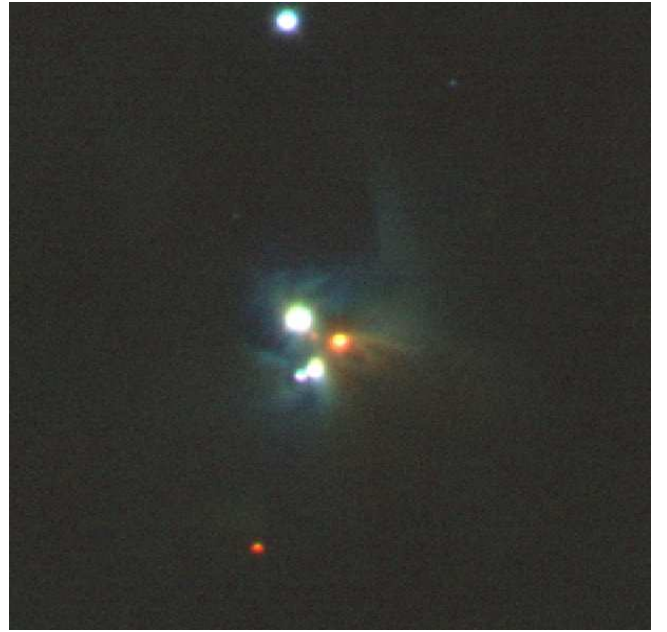}
\caption{False-colour image, composed from the \jb\ (blue), \h\ (green), and \textit{K\/}$_\mathrm{c}$ LIRIS images, observed on 2017 October 10. The displayed area is about $1.5\times1.5$~arcmin. North is up, and east is to the left.}
\label{Fig_lirim}
\end{figure}

\paragraph*{WHT LIRIS 2017}
We obtained near-infrared images and long-slit spectra of 2MASS~2235235+7517076 on 2017 October 10 with the LIRIS instrument installed on the 4.2-m William Herschel Telescope at the Observatorio del Roque de Los Muchachos (Spain). The images were taken in a 5-point dither pattern, through broad-band \jb,  \textit{H\/}, and narrow-band \textit{K\/}$_{c}$ filters, with total exposure times of 420.0\,s,  18.0\,s, and 60.0\,s, respectively. We took low resolution spectra in the \textit{ZJ\/} band, using the 0.75\,arcsec\ slit width, which yielded a spectral resolution of R=550--700 in the 0.9--1.4\,\micron\ range. Medium-resolution spectrum was obtained in the \textit{K\/}-band with the  1\,arcsec slit width, resulting in a spectral resolution of R=2500 in the 2.05--2.41\,\micron\ range. The measurements were performed with an ABBA nodding pattern. The total exposure times were 180\,s in the \textit{ZJ\/} and 60\,s in the \textit{K\/} band. For telluric correction and flux calibration we observed the A0-type star HIP~25357. The data reduction was done in the same way as in \citet{acosta2007}. The resulting \textit{K\/}-band spectrum is plotted in the lower panel of  Fig.~\ref{Fig_kspec}. A false-colour image, composed of the \jb, \h, and \textit{K}$_\mathrm{c}$ LIRIS images is shown in Fig.~\ref{Fig_lirim}.

\begin{table*}
\caption{\jhks\ fluxes available for 2MASS~22352345+7517076.}
\label{Tab_nir}
\begin{center}
\begin{tabular}{lcccccc}
\hline
\noalign{\smallskip}
Date of Obs. & MJD & F$_{J}$ &  F$_{H}$ & F$_{K_\mathrm{s}}$ & Telescope/ & Ref. \\ 
yyyymmdd &   & (mJy) & (mJy)  & (mJy) & Instrument \\
\noalign{\smallskip} \hline
 \noalign{\smallskip}
 19930830 & 49229 & 0.172$\pm$0.003 & $\cdots$ & 4.472$\pm$0.170 & CFHT/Redeye & 1 \\
 19991011 & 51462 & $\cdots$ & $\cdots$  & 15.415$\pm$0.760 & 2MASS & 2 \\
 20110919 & 55823 & $\cdots$ & $\cdots$ &  554.54$\pm$25.54 & CFHT/WIRCam & 0 \\ 
 20130826 & 56530 & 1.51$\pm$0.35 & 56.48$\pm$7.15 & 596.39$\pm$69.36 & Okayama 1.8-m/ISLE & 3 \\
 20140711 & 56849 & $\cdots$ & $\cdots$ & 496.51$\pm$45.79 & CFHT/WIRCam & 0 \\ 
 20160812 & 57612 & 0.341$\pm$0.009 & 10.566$\pm$0.29 & 189.99$\pm$9.10 & NOT/NOTCam & 0  \\ 
 20170728 & 57962 & 0.27$\pm$0.01 & 7.15$\pm$0.33 & 103.64$\pm$6.68  & NOT/NOTCam & 0  \\
 20171010 & 58036 & 0.31$\pm$0.01 & 8.45$\pm$0.37 & 151.75$\pm$8.25 & WHT/LIRIS & 0  \\
\hline
\end{tabular}
\end{center}
\flushleft{\small
References: 0--present work; 1--\citet{RD95}; 2--\citet{2MASS}; 3--\citet{onozato2015};}
\end{table*}

\subsection{Contribution of neighbouring sources to the mid-infrared fluxes}
\label{Sect_fluxcor}

2MASS \ks-band and \spitzer\ IRAC photometric data are available for the three brightest sources RD95~A, B, and D (IRS\,3, IRS\,2, and IRS\,1 in table~3 in \citet{kim2015}, respectively), associated with IRAS~22343+7501. To estimate the contribution of  IRS\,2 and IRS\,3 to the mid-infrared fluxes measured by \iras, \akari, \spitzer\ MIPS, and \wise, we extrapolated their SED slopes, measured on the 3.6--8\,\micron\ region, to 25\,\micron. We excluded the \ks-band flux to avoid effects of variability and extinction. 
The spectral indices $d\log(\lambda F(\lambda)/d\log(\lambda)$ over the IRAC wavelength interval of IRS\,2 and IRS\,3 are $-0.35$ and $-1.45$, respectively. Comparison of our near-infrared observations in 2016 and 2017 with the 2MASS data has shown that these stars did not exhibit large-amplitude variations during the period of observations. We assume that the contributions of these stars to the total fluxes beyond 25\,\micron, as well as the contribution of the much fainter RD95\,C and RD95\,E sources are negligible, and the \wise\ \textit{W1\/} and \textit{W2\/} fluxes of IRS\,1 and IRS\,2 are identical with their IRAC 3.6 and 4.5\,\micron\ fluxes. 
The fluxes of both sources, estimated for the \iras, \akari, \spitzer\ MIPS, and \wise\ bands, and their uncertainties, derived from the errors of the spectral indices, are shown in Table~\ref{Tab_minus}. The last column shows the fluxes of RD95\,D (2MASS~22352345+7517076), after subtracting the estimated fluxes of both neighbours from the catalogued data. Photometric data for $\lambda > 25$\,\micron\ are listed chronologically in Table~\ref{Tab_fir}.
 
 \begin{table*}
 \caption{Estimated mid-infrared fluxes of RD95\,A, RD95\,B, and RD95\,D (2MASS~22352497+7517113, 22352442+7517037, and 22352345+7517076, respectively)}
 \label{Tab_minus}
 \begin{center}
 \begin{tabular}{llccc}
 \hline
 \noalign{\smallskip}
 Band  & Date of Obs. & RD95\,A & RD95\,B & RD95\,D  \\
    & (Jy) & (Jy) & (Jy) &  \\
 \noalign{\smallskip}
 \hline
 \wise\ W1 & 2010 Feb 05  & 0.532$\pm$0.022 & 0.183$\pm$0.003 & 5.83$\pm$0.93	  \\
 \wise\ W2 & 2010 Feb 05  & 0.610$\pm$0.011 & 0.114$\pm$0.004 & 28.8$\pm$14.53      \\
 \wise\ W3 & 2010 Feb 05  & 1.16$\pm$0.11 & 0.11$\pm$0.01 & 37.31$\pm$3.37	     \\
 \wise\ W4 & 2010 Feb 05  & 1.71$\pm$0.24 & 0.08$\pm$0.005 & 68.22$\pm$1.20	    \\
 \spitzer\ MIPS 24\,\micron\ & 2004 Sep 24 & 1.79$\pm$0.27 & 0.08$\pm$0.005 & 22.75$\pm$1.00  \\
 \akari\ IRC 9\,\micron  & 2006 & 0.96$\pm$0.07 & 0.12$\pm$0.01 & 15.64$\pm$3.12      \\
 \akari\ IRC 18\,\micron & 2006 & 1.50$\pm$0.19 & 0.09$\pm$0.005 & 32.22$\pm$4.46  \\
 \iras\ 12\,\micron & 1983 & 1.16$\pm$0.11 & 0.11$\pm$0.002 & 4.43$\pm$0.02  \\
 \iras\ 25\,\micron & 1983 & 1.86$\pm$0.28 & 0.08$\pm$0.005 & 25.6$\pm$1.20  \\
\hline
 \end{tabular}
 \end{center}
 \end{table*}

\begin{table*}
\caption{Photometric data for wavelengths $\lambda \ge 60$~$\mu$m}
\label{Tab_fir}
\begin{center}
\begin{tabular}{lccccc}
\hline
\noalign{\smallskip}
Date of Obs. & Wavelength & Flux & eFlux & Telescope/Instrument & Ref. \\ 
    &   ($\mu$m)  & (Jy) & (Jy) \\
\noalign{\smallskip}
\hline
\noalign{\smallskip}
1983	     & 60     &  61.1	 & 2.5   & \iras \\
1983	     & 100    & 77.9	 & 3.1   & \iras \\
1993 Aug 9   & 800    & 0.710    & 0.114 & JCMT/UKT14  & (1) \\
1993 Aug 9   & 1100   & 0.383    & 0.030 & JCMT/UKT14  & (1) \\
1993 Aug 9   & 1300   & 0.232    & 0.024 & JCMT/UKT14  & (1) \\
1996 Dec 31  & 100    &   75.5   & 5.41  & \iso/ISOPHOT & (0) \\
1996 Dec 31  & 120    &   71.8   & 5.09  & \iso/ISOPHOT & (0) \\
1996 Dec 31  & 200    &  105.4   & 8.48  & \iso/ISOPHOT & (0) \\
2003 Sep     & 350    & 11.80	 & $\cdots$ & SHARC-II & (4) \\
2004 Sep 24  & 70     &  57.800  & 5.744 & \spitzer/MIPS & (0) \\
2007	     & 65     &  69.604  & 0.256 & \akari/FIS & (2) \\	  
2007	     & 90     &  84.563  & 0.042 & \akari/FIS &(2) \\  
2007	     & 140    & 118.452  & 0.158 & \akari/FIS &(2) \\   
2007	     & 160    & 125.983  & 0.484 &  \akari/FIS &(2) \\ 
2009 Dec 28--2010 Jan 25 &  70    &  96.248  & 0.739 & Herschel/PACS & (3) \\
2011 Jun 29  & 100    &  82.395  & 1.309 & \herschel/PACS & (3)\\ 
2011 Jun 29  & 160    &  81.168  & 4.451 & \herschel/PACS & (3)\\
2012--2014   & 450    & 14.62	 & $\cdots$ & JCMT/SCUBA-2 & (5) \\
2012--2014   & 850    & 2.18	 & $\cdots$  & JCMT/SCUBA-2 & (5) \\
2007 Oct 17  & 1300   & 0.03523  & $\cdots$  & SMA & (6) \\
\hline
\end{tabular}
\end{center}
\flushleft{\small
References: 0--present work; 1--\citet{RD95}; 2--FIS BSC Version~2, \citet{yamamura2016}; 3--\citet{Marton2017}; 4--\citet{suresh2016}; 5--\citet{Pattle2017}; 6--\citet{kim2015}}
\end{table*}

\section{Results} 
\label{Sect_res}

\subsection{The environment of the outbursting star}

\begin{figure}
\centering \includegraphics[width=\columnwidth]{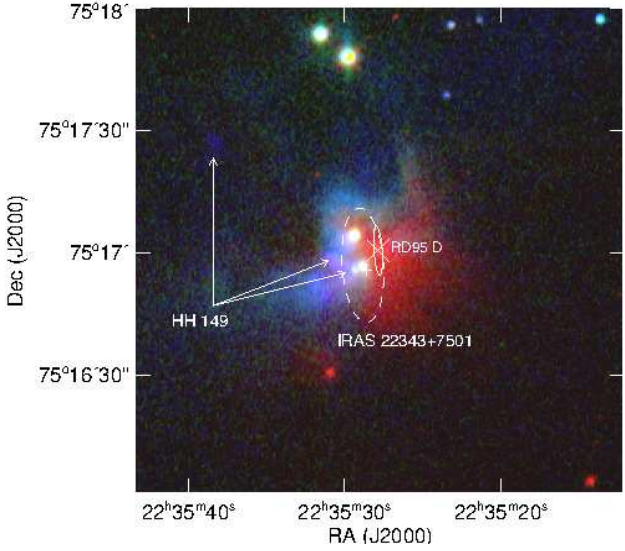}
\caption{False-colour image of IRAS~22343+7501, composed from the \ks-band WIRCam image obtained in 2011 (red), \ic-band image obtained in 2009 (green), and narrow-band \ion{S}{ii} image observed in 1990 (blue). Positions of the HH~149 knots \citep{balazs1992} are indicated by arrows. Dashed ellipse indicates the uncertainty ellipse of IRAS~22343+7501 from the \textit{IRAS Point Source Catalog\/}, and the ellipse drawn by solid line shows the position of the same source in the \textit{IRAS Faint Source Catalog\/}. The red source in the middle, marked by a white `x', is the outbursting star RD95\,D~$\equiv$~2MASS~22352345+7517076.}
\label{Fig_rgb}
\end{figure}

The young stars of the small \iras~22343+7501 cluster illuminate the optical reflection nebula RNO\,144, excite the Herbig--Haro jet HH\,149, and drive several molecular outflows \citep{sato1989,nikolic2003,kim2015}. To explore the morphology of the region we show in Fig.~\ref{Fig_rgb} a false-colour image, composed from post-outburst \ks, \ic, and pre-outburst narrow-band \ion{S}{ii} images. The uncertainty ellipses of the \iras\ source, listed in both the \iras\ \textit{Point Source Catalog} and  \textit{Faint Source Catalog\/} are overplotted, and position of the outbursting star 2MASS~22352345+7517076, as well as and the  knots of HH\,149 are marked. Figure~\ref{Fig_rgb} suggests that the driving source of HH\,149 probably is not RD95\,D, but other, unidentified member(s) of the \iras~22343+7501 cluster. Possibly different knots have different sources of excitation. Comparison of our \ic-band images obtained in 1999 and 2009 (Fig.~\ref{Fig_images}) suggests that the shape and brightness of the optical nebula was not significantly affected by the outburst. The bright infrared nebula surrounding the outbursting star on the south-western side probably is a result of the outburst. 

\subsection{Foreground extinction toward the line of sight of IRAS~22343+7501}
\label{Sect_ext}

The total extinction of a protostar consists of an interstellar and a circumstellar component. Although the central protostar is dimmed and reddened by the sum of both components, the circumstellar dust reradiates the absorbed light in the far-infrared, and thus contributes to the total luminosity of the system. The post-outburst \jb$-$\h\ vs. \h$-$\ks\ colour--colour diagram of the outbursting star, displayed in the left panel of Fig.~\ref{Fig_colour}, suggests a total extinction of $A_\mathrm{V} \approx 32$\,mag, whereas other members of the L1251\,C clump, whose 2MASS colours are plotted in the same diagram, have $A_\mathrm{V} \loa 10$~mag.
Based on spectral classification and optical photometry \citet{kun2009} derived foreground extinctions of \av=8.88 and 9.88\,mag for RD95\,A and RD95\,B, respectively. We adopt the average, \av =9.4\,mag, for the extinction originating from the molecular cloud clump, embedding these three stars. It coincides with the result in \citet{RD95}, but is larger than the \av=5.4\,mag, derived by \citet{Dunham2013}, by averaging the extinction of Class~II sources over a wide area of the Cepheus flare star forming region.

\subsection{Brightness and colour evolution} \label{Sect_lc}

\begin{figure}
\centering \includegraphics[width=\columnwidth]{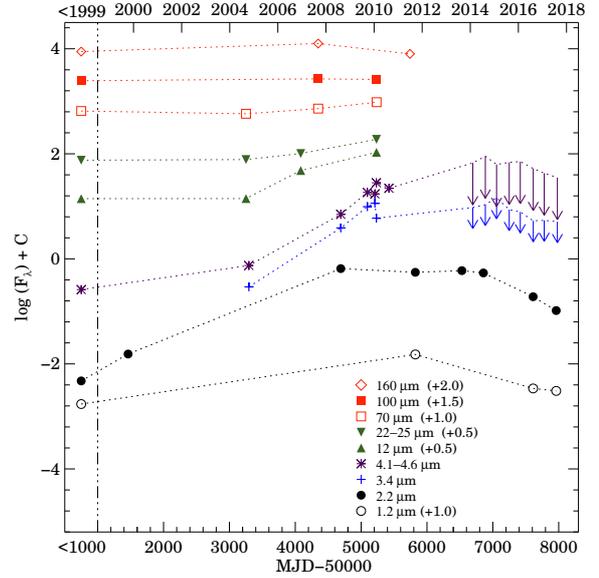}
\caption{Flux evolution of 2MASS~22352345+7517076. Logarithms of flux densities measured in Jy, and shifted as indicated in parentheses in the lower right corner are plotted. Arrows indicate dates and flux ranges of \textit{NEOWISE\/} measurements.}
\label{Fig_lc}
\end{figure}

Figure~\ref{Fig_lc} shows a multi-wavelength light curve of 2MASS~22352345+7517076 between 1983 and 2017. Errorbars are smaller than symbol sizes. Data points on the left side of the dash-three-dots line, that is earlier than 1999, were actually measured at different epochs. We regard them as quiescent phase fluxes. Fluxes at 12\,\micron\ were measured by \textit{IRAS\/} and \wise. For \spitzer\ and \akari\ we interpolated the flux logarithms between 8 and 24\,\micron, and 9 and 18\,\micron, respectively, to obtain 12-\micron\ data. We indicated with arrows dates and flux ranges of \textit{NEOWISE\/} observations. At 2-\micron\ the flux tripled between 1993 and 1999, and increased by a factor of 40 between 1999 and 2008. It had a six-year long plateau between $\sim 2008$ (\akari\ post-helium) and $\sim 2014$, and then dropped by a factor of 2.8 between 2014 July and 2016 August. Our NOTCam and LIRIS data, obtained in 2017, show further fading. The latest 2-\micron\ flux was still more than seven times higher than the 2MASS level. Comparison of the ISOCAM and \textit{Spitzer c2d\/} 4.5-micron data indicates brightening of the source between 1997 and 2004.  In the 3.6--4.5\,\micron\ region the flux rose steeply between 2004 and 2009. The 12 and 24-\micron\ flux stayed constant between the \textit{IRAS\/} and \textit{Spitzer c2d\/}. The 60-\micron\ flux slightly increased between 2004 and 2010, whereas the 100-\micron\ fluxes, measured by various instruments over the 1983--2010 interval, well coincide with each other. At 160~\micron\ the slight descending may result from the different angular resolutions of the \akari\ FIS and \herschel\ PACS instruments. 

\begin{figure}
\centerline
{\includegraphics[width=0.25\textwidth]{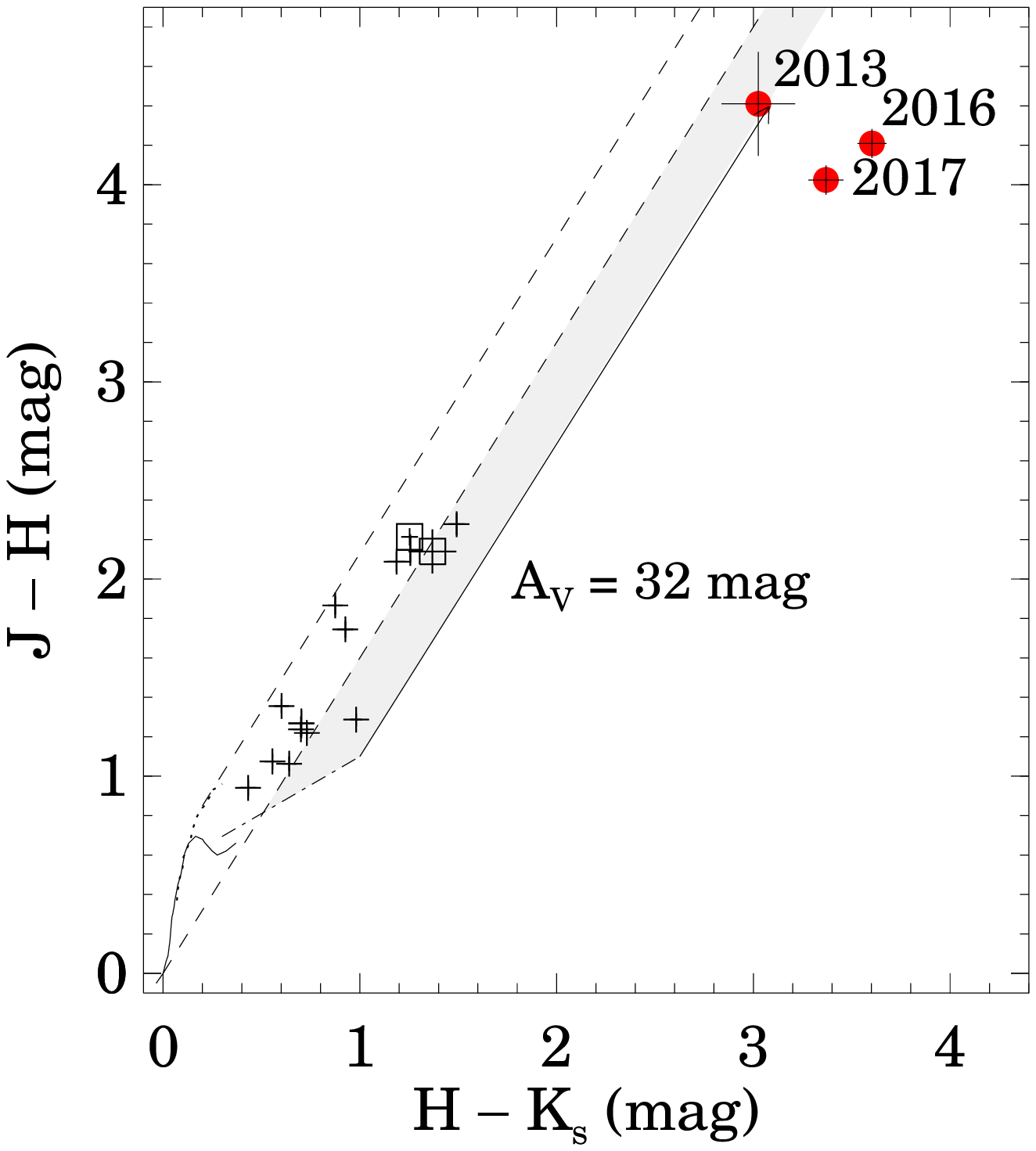}\includegraphics[width=0.25\textwidth]{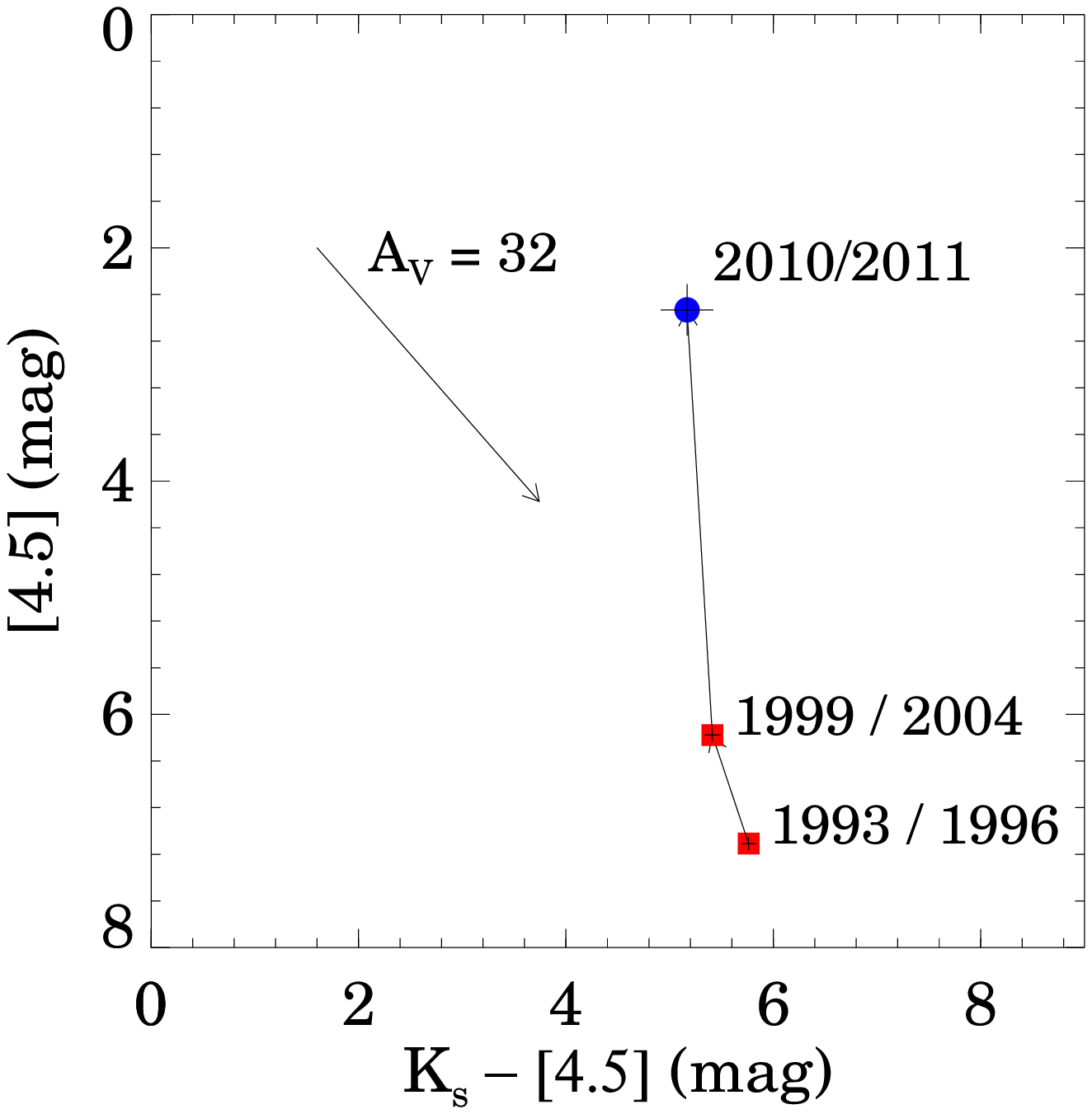}}
\caption{\textit{Left:\/} $J-H$ vs. $H-K_\mathrm{s}$ colour--colour diagram of young stellar objects associated with the molecular cloud clump L1251\,C. Red filled circles show the colours of 2MASS~22352345+7517076 measured at three post-outburst epochs. Squares indicate the 2MASS colours of RD95\,A and RD95\,B, and small crosses mark  other YSOs identified in the \textit{Spitzer c2d\/} data \citep{kim2015}. The normal main sequence (solid line), the band of the reddened main sequence (dashed lines), and the T~Tauri locus \citep*[dash-dotted line,][]{Meyer} are also drawn. The shaded area indicates the location of reddened \textit{K}$_s$-band-excess pre-main-sequence stars. \textit{Right:\/} $[4.6]$ vs. \ks$-[4.6]$ colour--magnitude diagram of 2MASS~22352345+7517076 at four epochs of the outburst (blue 
symbols) and at the low and intermediate states before 2000 (red squares). The arrow indicates the effect of an extinction of \av=32\,mag in each diagram.}
\label{Fig_colour}
\end{figure}

For examining colour variations associated with the outburst 1.2--4.6~\micron\ data are available. The \jb$-$\ks\ colour index increased from 4.55 to 7.43 while the \ks\ magnitude brightened from 12.87 to 7.62 \citep{RD95,onozato2015}. This {\em redder when brighter} behaviour is rare among the eruptive YSOs examined by \citet{antoniucci2014}, and suggests variations in the disc. It shows that the proportion of scattered light is higher in the low-state \jb\ flux. The right panel of Fig.~\ref{Fig_colour} shows a \ks\ vs. (\ks$-[4.5]$) colour--magnitude diagram for three brightness stages. The strong brightening without colour change suggests the appearance of a hot source in the system between 2004 and 2010.

\subsection{Near-infrared spectra}

The first spectrum of IRAS~22343+7501, displayed in Fig.~\ref{Fig_ircsp}, was detected by the \akari\ IRC on 2008 August 12, on the rising part of the light curve. The flux calibration of this spectrum has been lost due to the saturation, thus we are restricted to qualitative statements. The slope of the spectrum confirms the high extinction of the source. The conspicuous absorption bands of the H$_2$O ice at 3\,\micron\ and that of the CO$_2$ ice at 4.27\,\micron\  resemble the spectra of edge-on YSO discs, observed with the same instrument \citep{aikawa2012}. No accretion signature can be seen, although the observed wavelength region contains several hydrogen lines, common in YSO spectra \citep{beck2007}.

The featureless \textit{K\/}-band spectrum observed in 2016 August, shown by the black curve in the upper panel of Fig.~\ref{Fig_kspec}, resembles the spectrum of the object observed in 2013 by \citet{onozato2015}. Neither photospheric absorption lines nor accretion-indicator emission lines can be seen. The absence of atomic lines indicates strong veiling due to the hot gas in the innermost part of the circumstellar region. The low signal-to-noise spectrum obtained with the same instrument on 2017 July 28 (shown by blue line in the upper panel of Fig.~\ref{Fig_kspec}) suggested the possible presence of emission lines. Emission lines of the molecular hydrogen at 2.12, 2.22, and 2.40~\micron\ are clearly visible in the most recent spectrum, observed in 2017 October. These lines originate from shocks, due to strong accretion-related winds. Accretion-related winds expand with characteristic speeds of 100--500\,km\,s$^{-1}$. If the outflow, giving rise to the H$_2$ lines detected in 2017, started some ten years earlier, near the onset of the outburst, then the shocked gas, emitting the lines, is located at some 300--1000\,AU from the wind launching region, around the outer edge of the outflow cavity. The available spectra display none of the FUor characteristics established by \citet{Connelley2018}.

\subsection{Spectral energy distribution} \label{Sect_sed}

\begin{figure}
\centering \includegraphics[width=\columnwidth]{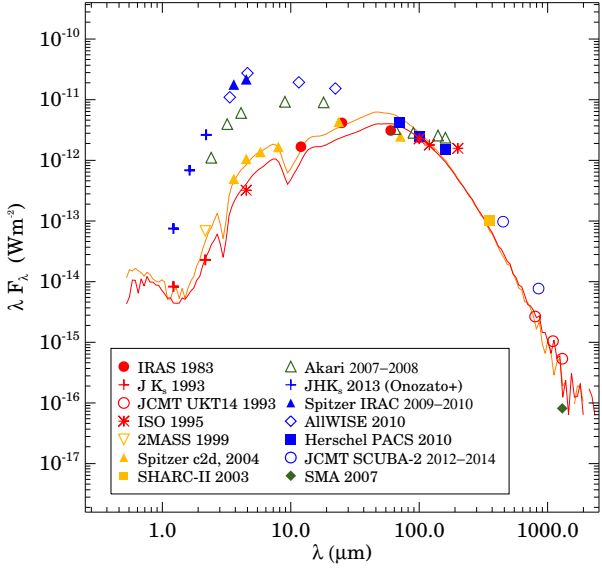}
\caption{Spectral energy distribution of 2MASS~22352345+7517076 at various epochs,
corrected for a foreground extinction of 9.4~mag. Red symbols show data from the period 1993--1996, orange symbols indicate fluxes measured between 1999 and 2004, green symbols mark \akari\ data from 2006--2008, and data obtained between 2010 and 2014 are plotted with blue symbols. }
\label{Fig_sed1}
\end{figure}

We constructed the SED of 2MASS~22352345+7517076 using all data listed in Sect.~\ref{Sect_data}. The mid-infrared fluxes were corrected for the contribution of neighbouring sources. 
Figure~\ref{Fig_sed1} shows the SED corrected for the adopted foreground interstellar extinction of $A_\mathrm{V}=9.4$~mag. Wavelength dependence of extinction was adopted from \citet{cardelli1989} with $R_{V}=5.5$ for the near-infrared, and from \citet{xue2016} for longer wavelengths. Different symbol colours indicate different epochs. Actually an {\em epoch} may cover more than one year, nevertheless they represent specific brightness levels. The spectral index $\alpha=d\log(\lambda F(\lambda))/d\log(\lambda)$ of the extinction-corrected pre-outburst SED over the 2.0--25\,\micron\ interval, defined by the \citet{RD95} \textit{K}$^\prime$-band and \iras\ 25-\micron\ flux, is $\alpha_{low}=1.69$, characteristic of a Class~I YSO \citep{greene94}. The same index of the outburst SED, defined by the $K_\mathrm{s}$ measurement by \citet{onozato2015} and \wise\ 22-\micron\ flux, is $\alpha_{high}=0.76$, also indicative of a Class~I source. 

We determined bolometric temperatures and luminosities \citep{Myers1993} by integrating the SED, corrected for the adopted interstellar component of extinction, for each epoch. The results are listed in Table~\ref{Tab_sed}. The \textit{T\/}$_\mathrm{bol}$ values are also indicative of a Class~I YSO.

 \begin{table}
 \caption{Bolometric temperatures and luminosities derived for various epochs.}
 \label{Tab_sed}
 \begin{center}
 \begin{tabular}{lccr}
 \hline
 \noalign{\smallskip}
 Epoch & Period & \textit{T\/}$_\mathrm{bol}$ & \textit{L\/}$_\mathrm{bol}$  \\
 \noalign{\smallskip}
 \hline
 & & (K) & (L$_{\sun}$) \\
 \noalign{\smallskip}
 \hline
1 & 1983--1996 & 138.0 & 32.2  \\
2 & 1999--2004 & 177.0 & 32.7  \\
3 & 2007--2008 & 410.0 & 85.0  \\
4 & 2010--2013 & 483.0 & 165.0  \\
 \hline
 \end{tabular}
 \end{center}
 \end{table}

We compared the extinction-corrected low-state SED with two-dimensional protostellar models, using the 2008 version of the \citet{whitney2003} radiative transfer code. The model systems consist of a central star, a circumstellar disc and an infalling envelope. The envelope contains an outflow cavity of  polynomial-shape. This procedure allows us to assess whether the structure of our target is consistent with such a protostellar model or not, and obtain an insight into the possible range of physical parameters of the central star and the envelope. We used the dust files distributed with the 2008 version of the code. Some initial parameters can be constrained by the shape of the SED.

The low-state bolometric luminosity is composed of the luminosity of the central star and disc accretion luminosity \citep{white2004}. We can constrain the luminosity of the central object with the assumption that \lbol\ in quiescence is dominated by the contribution of the central star. It is fulfilled if the disc accretion rate is lower than some $10^{-6}$\msunyr. \lbol=32\lsun\ suggests a central protostar of 1.6--1.8\msun\ near the $10^{5}$-yr isochrone of \citet*{siess2000}. We tentatively adopt this mass and age, which imply a stellar radius of 9\,R$_{\sun}$. The density of the envelope, a function of the infall rate, determines the circumstellar extinction and shapes the SED at the shortest and longest wavelengths \citep[cf.][]{furlan2016}. Absence of conspicuous silicate absorption at 10\,\micron\ suggests a moderate inclination ($i \loa 45\degr$). We set the envelope outer radius to 10000~AU, a typical size for low-mass protostars. We found that $1.5\times10^{-5}$\msunyr $\la \dot{M}_\mathrm{infall} \la 3.0\times10^{-5}$\msunyr\ is needed to approximate the submillimeter data points satisfactorily. Furthermore, though the shape of the model SED is less sensitive to disc mass, we set it to 0.024\msun, suggested by the SMA 1.3-mm measurement \citep{kim2015}. We set the disc outer radius to 200\,AU, and assume that the inner radius of the dust disc in quiescence is at the dust sublimation radius. Various combinations of disc accretion rate ($5.0\times10^{-8} \la \dot{M}_{acc}/M_{\sun}$\,yr$^{-1} \la 9.0\times10^{-7}$), outflow cavity opening angle ($10\degr \loa \theta \loa 30\degr$), and inclination ($\theta \loa i \loa 45\degr$) result in satisfactory fitting to the quiescence SED. Keeping in mind the inherent degeneracies in the modelling and the non-simultaneity of our SED data we do not intend to find a mathemathically best-fitting model. The goodness of a model was judged by eye, looking at its apparent compatibility with the observed SED.
 
We find that this simple protostellar model is able to reproduce the observed low-state SED. The estimated infall rate of $\sim2.0\times10^{-5}$\msunyr\ is consistent with the adopted mass and age of the central star.
The red solid line in Fig.~\ref{Fig_sed1} shows a model that fits well the low-state (Epoch~1) data. The disc accretion rate of the model is $\dot{M}=3.5\times10^{-7}$\msunyr, the inner radius of the disc is 1.0\,r$_\mathrm{subl}$, and the cavity opening angle is 20\degr. The inclination was set to 27$\degr$. The total luminosity of the model, 25.6\lsun, is consistent with \lbol, taking into account the contribution of scattered and reprocessed radiation to \lbol, due to the low inclination \citep{whitney2003}. The envelope mass of the model is 1.13\msun, compatible with the 1.19--1.67\msun, resulted from the SCUBA--2 850-\micron\ data \citep{Pattle2017}. 
The orange line shows a model SED, derived from the low-state one by increasing $\dot{M}_{acc}$ to $2\times10^{-6}$\msunyr, which fits satisfactorily the Epoch~2 data.

The SED of Epoch~3, defined by \akari\ data between 2.4 and 160\,\micron, cannot be fitted with a protostellar model, suggesting that the accretion rate is is no longer constant across the whole disc, and the major source of radiation is not the central protostar, but a hot, luminous accretion disc. 

\subsection{Accretion disc modelling}
\label{Sec_accdisc}

In order to study the accretion rate variations in a quantitative way, and to
try to separate the effects of changing extinction and accretion rate, we
fitted the near-infrared part of the SED of each epoch using a simple accretion disc model. Following
our successful approach modelling  the near-infrared SEDs of HBC~722
\citep{kospal2016},  V346~Nor \citep{kospal2017a}, and V582~Aur
\citep{abraham2018}, we adopted a steady optically thick and geometrically
thin viscous accretion disc, with a radially constant mass-accretion rate (see
Eq.~ 1 in \citet{kospal2016}). The synthetic disc SEDs were calculated by
integrating the blackbody emission of concentric annuli starting from the
stellar radius out to $R_{out}$. The outer radius was fixed to $R_{out}$ = 2~AU (the exact value did not affect the results), thus we are left with only
two free parameters: the product of the stellar mass and the accretion rate,
and the line-of-sight extinction \av. We fixed the stellar mass and radius to 1.6~M$_{\odot}$ and 9.0~$R_{\odot}$ (Sect.~\ref{Sect_sed}), respectively. For the disc inclination 27$\degr$ was taken. We assumed that the observations obtained between 1993 and 1996 well represent the pre-outburst state, thus a quiescent SED was compiled from these data, and added to our synthetic accretion disc SED. The resulting fluxes were then reddened using a large grid of \av\ values and the standard extinction law from \citet{cardelli1989} with $R_\mathrm{V}$ = 5.5. The fitting procedure was performed with ${\chi}^2$ minimization, and the formal uncertainties of the fitted parameters (\av\ and $\dot{M}_{acc}$) were computed with a Monte-Carlo approach.

\begin{figure*}
\centering \includegraphics{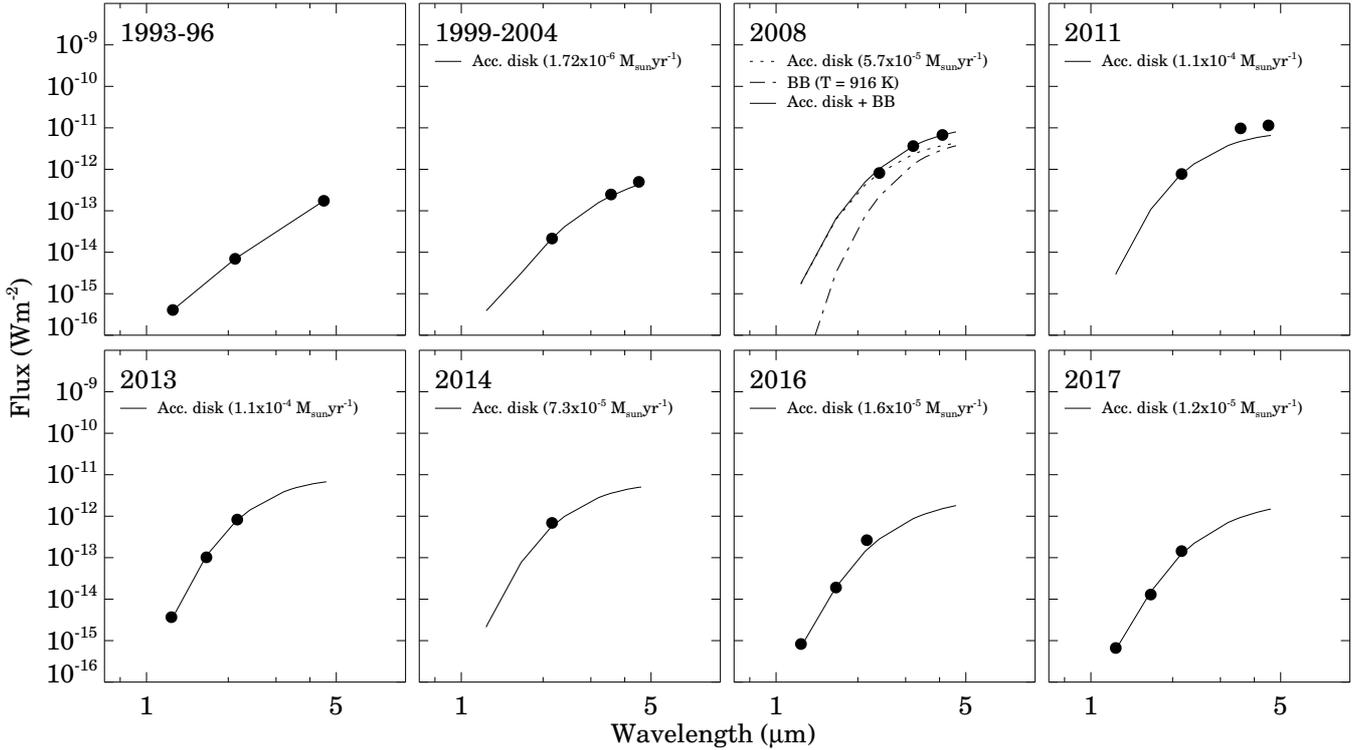}
\caption{SEDs at eight different epochs of the outburst. Each panel presents
the actual measurements (filled circles). The first epoch in 1993-96
corresponds to the quiescent phase. Overplotted are the accretion disc fits to
\jhks\ data points, as well as a blackbody fit to the mid-infrared excess in 2008.}
\label{fig_modelSED}
\end{figure*}

Since the data at the mid-infrared wavelengths may be contaminated by thermal
emission of the dust disc, in the first step we fitted only those epochs when
\jhks\ data points were available. The resulting three independent
values for \av\ ($31.5\pm1.5$ in 2013; 33.0$\pm$0.8 in 2016; and 32.07$\pm$1.02 in 2017) were consistent within their error bars. Since in Sect.~\ref{Sect_ext} we presented additional arguments that the extinction is practically invariable, we fixed  \av=32.19~mag, and repeated our modelling fitting only the accretion rate as the only free parameter. With this modification we were able to fit the SEDs at all the remaining epochs, because already a single K-band magnitude was sufficient to determine $\dot{M}_{acc}$. The near-infrared SEDs and the fitted accretion disc models are presented in Fig.~\ref{fig_modelSED}.

When all \jhks\ data are available, our models reproduce them well. At the earlier epochs (before 2008) a pure accretion disc can eventually fit the mid-infrared points as well. 
A definite excess over the disc model can be seen in the \akari\ data (2008). 
In order to characterize this excess, we reproduced the 3.2 and 4.1\,\micron\ excess flux values with a Planck-function, whose temperature is also shown in the figure panel.

\begin{figure}
\centering \includegraphics[width=\columnwidth]{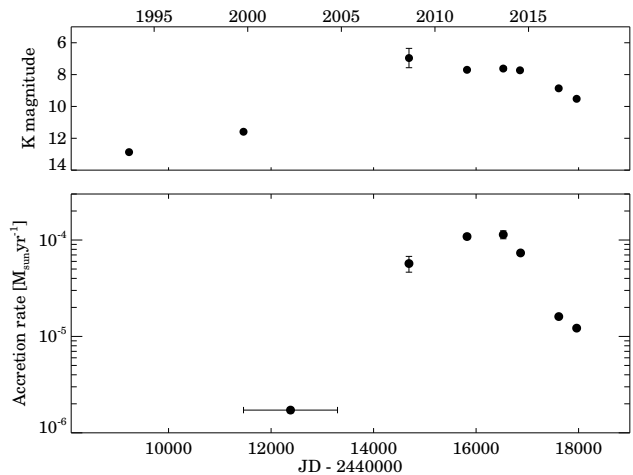}
\caption{Lower panel: Temporal evolution of the accretion rate derived from
our simple accretion disc model fitted to the near-infrared spectral energy
distribution (Sect.~\ref{Sec_accdisc}). The line-of-sight extinction was
assumed to be constant (\av=32.19\,mag) during the whole outburst. The upper
panel shows the \ks-band light curve for reference.}
\label{fig_accdisc}
\end{figure}

In Fig.~\ref{fig_accdisc} we plotted the resulting $\dot{M}_{acc}$ values as a function
of time. Associated with the brightness maximum between 2010 and 2015 a
definite peak is visible in the accretion rate. While the quiescent accretion
rate is unknown, between 2003 and 2010 an $\dot{M}_{acc}$ increase of about two orders
of magnitude occurred. Following a few years at peak accretion rate a rapid
decline started around 2015, which may indicate the end of the outburst. The
maximum value provided by our model was about 10$^{-4}$\msunyr,
a typical values for FU~Ori-type outbursts \citep{audard2014}.

\section{Discussion}
\label{Sect_disc}

Our results suggest that 2MASS~22352345+7517076 has recently undergone a powerful accretion burst. Both the change in the derived bolometric luminosity and the accretion disc modelling suggest accretion rates about 10$^{-4}$\msunyr, typical of FUor outbursts. The major infrared space missions \iras, \iso, \spitzer, \akari, \wise, and \herschel\ observed specific stages of the outburst over a wide wavelength interval. The available archival infrared data, spanning a 35-years interval, allowed us to constrain some properties of the central star and some details of the outburst process.

\subsection{The central protostar}

Our target is probably not older than 1--2$\times10^5$~years, the typical age of Class~I YSOs \citep{Kristensen2018}. Its luminosity of 25\lsun, estimated from the low-state \lbol, suggests a central protostar of 1.6--2.0\msun. It will arrive at the main sequence as a mid A--early F type star. We observe it through an extinction of \av $\approx 32$\,mag, which prevents us from detection of the stellar photosphere directly. Of the total 32\,mag, \av $\approx 9.4$\,mag arises from the foreground cloud and embedding molecular clump. 
The shape of the low-state SED suggests an order of magnitude higher rate of mass infall from the envelope than the disc accretion rate. This situation leads to episodic accretion bursts \citep{Bell1994,Bell1995}. 

\subsection{Evolution of the SED}

\begin{figure}
\centering \includegraphics[width=\columnwidth]{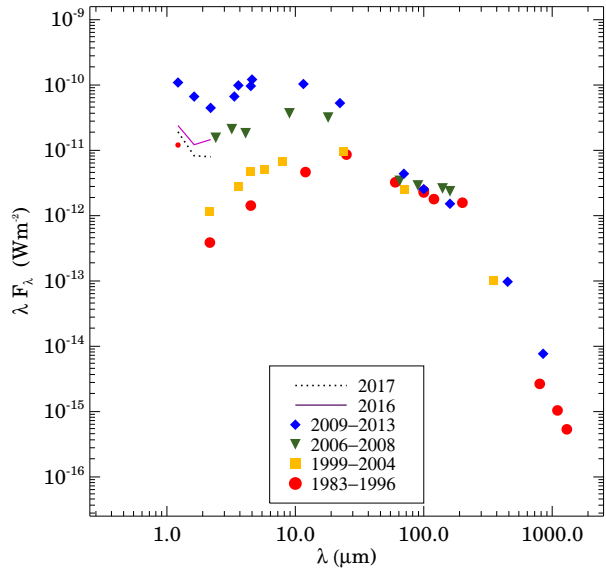}
\caption{Spectral energy distribution of 2MASS~22352345+7517076, corrected for the total extinction of \av=32~mag. }
\label{Fig_sed-ext}
\end{figure}

\paragraph*{Rising.} 
To get an insight into the variations of the central regions of the system we plotted in Fig.~\ref{Fig_sed-ext} the SEDs for different epochs, corrected for a total extinction of \av=32.19\,mag. We could see a SED like this looking at the system face-on, except the optical region, missing from our data set. Correcting the low-state \jb-band flux for the total extinction we obtain the unlikely high position plotted with smaller symbol, indicating that we observe scattered light, originated from outer regions of the disc atmosphere. Figure~\ref{Fig_sed-ext} suggests that the long-term rising of the near-infrared fluxes of 2MASS~22352345+7517076 consisted of two stages. The near-infrared fluxes, originating from the innermost regions of the circumstellar disc, increased between 1993 and 2004, whereas the 24-\micron\ flux measured by MIPS in 2004 was virtually same as the \iras\ 25-\micron\ flux in 1983. These data point to the process of mass accumulation in the inner disc region. During this phase the nearly three-fold flux increment at 2.2 and 4.5\,\micron\ was probably caused by a similar growth of the emission area due to expansion of the dust destruction front. The mass, piled up during the eight years between the \textit{ISOCAM\/} and \textit{Spitzer c2d\/} observations, might drifted from a distance of 7--8~AU to the dust destruction radius \citep[cf.][]{Kuffmeier2018}.
 
The \akari\ data in 2007 show a next stage of the outburst: the strong brightening in the mid-infrared and even a noticeable flux increase around 100\,\micron\ show that the radiation of hot inner disc started heating the envelope. Theoretical considerations \citep{Johnstone2013} have shown that the envelope responds quickly to heating in the mid-infrared, thus the actual outburst, the activation of magneto-rotational and thermal instability \citep{Zhu2009} occurred between 2004 and 2007. The near-infrared \akari\ data, obtained in 2008 August, suggest the emergence of a hot central region. The high extinction prevents us from detecting the hottest region of the system. Its emergence, however, is reflected by the increased mid-infrared fluxes. \spitzer, \wise, and \herschel\ detected further brightening from 2007 to 2010 over the 3.4--70~\micron\ interval. Comparison of the JCMT measurements from 1993 and 2014 shows an increment at submillimeter wavelengths. 

\paragraph*{Plateau phase} The outburst probably reached its brightness peak in 2009--early 2010.  Our optical, \ic-band images, obtained in 2009 October show a dim source, not detected earlier and too faint for photometry, at the position of 2MASS~22352345+7517076 (Fig.~\ref{Fig_images}). The same object, indicative of scattered light from the environment of the outbursting star, is also discernible in the \textit{Pan-STARRS} \citep{Chambers2016} \textit{y}-band stacked image, observed between 2010 and 2015 and available in the \textit{Pan-STARRS1\/} data archive\footnote{\url{https://panstarrs.stsci.edu/}}. 
The peak of the SED shifted from the low-state $\approx25$\,\micron\ to $\approx4.6$\,\micron\ due to the increased luminosity. This wavelength shift indicates that the temperature of the envelope photosphere \citep{Hartmann1998} increased from $\sim110$\,K to $\sim630$\,K, leading to profound changes in the envelope composition and structure. 

\paragraph*{Fading.} The light curve in Fig.~\ref{Fig_lc} shows that the \ks-band flux of the protostellar system started declining in 2015. 
The present fading may either indicate the end of the outburst or may be episodic. Temporary decreases in accretion rates were observed in several FU~Ori type stars, e.~g. in V1647~Ori \citep{aspin2009}, HBC~722 \citep{kospal2016}, V899~Mon \citep{Ninan2015}, and V346~Nor \citep{kospal2017b}. The observed flux evolution of 2MASS~22352345+7517076 suggests that a clump of matter, resulting from gravitational instability of the outer disc regions, was accreted onto the star. The few-years time scale of variations in the accretion rate suggests that the clump might have formed within 10~AU from the centre. Figure~\ref{fig_accdisc} suggests that the central star accreted nearly $10^{-3}$\msun, about a Jupiter mass during the present outburst. 

\paragraph*{Comparison with other outbursting protostars.} 
The \textit{K\/}-band spectra of 2MASS~22352345+7517076 differ from those of bona fide  FUors \citep{Connelley2018}, and classify this star a {\em peculiar\/} eruptive star with some FUor-like properties. Its peak \lbol $\approx 165$\lsun\ was somewhat higher than the median bolometric luminosity (99\lsun) of the sample examined by \citet{Connelley2018}.

The \textit{VISTA Variables in the Via Lactea (VVV)} survey of the Galactic mid-plane \citep{contreras1} resulted in the discovery of 70 eruptive Class~I protostars. Their \textit{K\/}-band amplitudes are mostly smaller than 3~mag. The observed outburst of our target was an exceptionally energetic event. Most of the eruptive stars of VVV show near-infrared spectra characteristic of known FUor or EXor type stars. No featureless spectrum similar to that of our target can be found in the spectroscopically confirmed VVV sample \citep{contreras2}. 
The six-year duration of the plateau phase of the outburst is longer than the average of four years, estimated by \citet{contreras1} for the outbursting protostars of the VVV survey, and much shorter than a typical FUor outburst. The featureless \textit{K\/}-band spectrum, the outburst time scale, and bolometric temperature resemble the deeply embedded outbursting protostar OO~Ser \citep{kospal2007}, although 2MASS~22352345+7517076 is more massive and had some five times higher peak luminosity. The similarity suggests that the outburst attributes depend more on the mass reservoir available in the envelope than on the mass of the central star. 

The long brightening at the wavelengths below $\sim$10\,\micron, without appreciable increasing of the inner disc temperature suggests that a considerable portion of the eruptive stars detected by near-infrared surveys may stay in the phase of pre-outburst mass accumulation. The increased mid-infrared fluxes indicate the real outburst of embedded YSOs.

\section{Summary}
\label{Sect_sum}
We present infrared observational data of the protostar 2MASS~22352345+7517076, spanning a 35-year period. During this period the star underwent a strong accretion burst. The SED of the system over the 2--160\,\micron\ region was sampled at three specific brightness phases: (1) quiescence in 1983--1996; (2) slowly rising (dust accumulation) phase in 1999--2004; (3) outburst 2007--2010. 
 
We found that 2MASS~22352345+7517076 is probably a 1.6--2.0\msun\ Class~I young star. Its bolometric luminosity increased from 32\lsun\ to 165\lsun\ between 1993 and 2010. Variation in the SED shape suggests that the amount of the dust near the sublimation radius increased between 1997 and 2004. The outburst occurred between 2004 and 2007, when a hot central object appeared in the system. The peak accretion rate was $1.1\times10^{-4}$\msunyr, typical of FU~Ori type outbursts. 
 
The time scales of the outburst support the scenario presented in \citet{Kuffmeier2018}: a clump of a Jupiter mass, formed by gravitational instability of the disc at some 7--8~AU from the centre was accreted onto the star during the orbital time. 

The \textit{K\/}-band spectra, observed during the plateau phase, were strongly veiled, and H$_2$ emission lines, indicative of a new outflow, appeared in the spectrum obtained in 2017. These spectra classify 2MASS~22352345+7517076 into the group of peculiar eruptive young stars, defined by \citet{Connelley2018}, different from classical FUors.

The Herbig--Haro jet HH~149 is probably driven by another protostar of the small cluster associated with IRAS~22343+7501.  

\section*{Acknowledgements}
We are grateful to Jochen Eisl\"offel for sending us the optical images
obtained in 1990 with the 3.5-m telescope of the Calar Alto Observatory. 
This research utilized observations with \textit{AKARI\/}, a JAXA project with the 
participation of ESA. This work also utilized observations made with the \textit{Spitzer Space Telescope}, which is operated by the Jet Propulsion Laboratory, California Institute of Technology under a contract with NASA. This research has made use of the NASA/ IPAC Infrared Science Archive, which is operated by the Jet Propulsion Laboratory, California Institute of Technology, under contract with the National Aeronautics and Space Administration. This work has made use of data from the European Space Agency (ESA) mission \gaia\ (\url{https://www.cosmos.esa.int/gaia}), processed by the \gaia\ Data Processing and Analysis Consortium (DPAC, \url{https://www.cosmos.esa.int/web/gaia/dpac/consortium}). Funding for the DPAC has been provided by national institutions, in particular the institutions
participating in the {\it Gaia} Multilateral Agreement.  
Our research has benefited from the VizieR catalogue access tool, CDS, Strasbourg, France. Financial support by the Momentum grant of the MTA CSFK Lend\"ulet disc Research Group, the Lend\"ulet grant LP2012-31 of the Hungarian Academy of Sciences is acknowledged.

\bibliographystyle{mnras}
\bibliography{iras22343}

\begin{thebibliography}{}
\makeatletter
\relax
\def\mn@urlcharsother{\let\do\@makeother \do\$\do\&\do\#\do\^\do\_\do\%\do\~}
\def\mn@doi{\begingroup\mn@urlcharsother \@ifnextchar [ {\mn@doi@}
  {\mn@doi@[]}}
\def\mn@doi@[#1]#2{\def\@tempa{#1}\ifx\@tempa\@empty \href
  {http://dx.doi.org/#2} {doi:#2}\else \href {http://dx.doi.org/#2} {#1}\fi
  \endgroup}
\def\mn@eprint#1#2{\mn@eprint@#1:#2::\@nil}
\def\mn@eprint@arXiv#1{\href {http://arxiv.org/abs/#1} {{\tt arXiv:#1}}}
\def\mn@eprint@dblp#1{\href {http://dblp.uni-trier.de/rec/bibtex/#1.xml}
  {dblp:#1}}
\def\mn@eprint@#1:#2:#3:#4\@nil{\def\@tempa {#1}\def\@tempb {#2}\def\@tempc
  {#3}\ifx \@tempc \@empty \let \@tempc \@tempb \let \@tempb \@tempa \fi \ifx
  \@tempb \@empty \def\@tempb {arXiv}\fi \@ifundefined
  {mn@eprint@\@tempb}{\@tempb:\@tempc}{\expandafter \expandafter \csname
  mn@eprint@\@tempb\endcsname \expandafter{\@tempc}}}

\bibitem[\protect\citeauthoryear{{{\'A}brah{\'a}m}, {Leinert}, {Burkert},
  {Henning}  \& {Lemke}}{{{\'A}brah{\'a}m} et~al.}{2000}]{abraham2000}
{{\'A}brah{\'a}m} P.,  {Leinert} C.,  {Burkert} A.,  {Henning} T.,   {Lemke}
  D.,  2000, \aap, \href {http://adsabs.harvard.edu/abs/2000A%26A...354..965A}
  {354, 965}

\bibitem[\protect\citeauthoryear{{{\'A}brah{\'a}m} et~al.,}{{{\'A}brah{\'a}m}
  et~al.}{2017}]{abraham2018}
{{\'A}brah{\'a}m} P.,  et~al., 2017, preprint, \href
  {http://adsabs.harvard.edu/abs/2017arXiv171204968A} {} (\mn@eprint {arXiv}
  {1712.04968})

\bibitem[\protect\citeauthoryear{{Acosta-Pulido} et~al.,}{{Acosta-Pulido}
  et~al.}{2007}]{acosta2007}
{Acosta-Pulido} J.~A.,  et~al., 2007, \mn@doi [\aj] {10.1086/512101}, \href
  {http://adsabs.harvard.edu/abs/2007AJ....133.2020A} {133, 2020}

\bibitem[\protect\citeauthoryear{{Aikawa} et~al.,}{{Aikawa}
  et~al.}{2012}]{aikawa2012}
{Aikawa} Y.,  et~al., 2012, \mn@doi [\aap] {10.1051/0004-6361/201015999}, \href
  {http://adsabs.harvard.edu/abs/2012A%26A...538A..57A} {538, A57}

\bibitem[\protect\citeauthoryear{{Antoniucci}, {Giannini}, {Li Causi}  \&
  {Lorenzetti}}{{Antoniucci} et~al.}{2014}]{antoniucci2014}
{Antoniucci} S.,  {Giannini} T.,  {Li Causi} G.,   {Lorenzetti} D.,  2014,
  \mn@doi [\apj] {10.1088/0004-637X/782/1/51}, \href
  {http://adsabs.harvard.edu/abs/2014ApJ...782...51A} {782, 51}

\bibitem[\protect\citeauthoryear{{Aspin} et~al.,}{{Aspin}
  et~al.}{2009}]{aspin2009}
{Aspin} C.,  et~al., 2009, \mn@doi [\apjl] {10.1088/0004-637X/692/2/L67}, \href
  {http://adsabs.harvard.edu/abs/2009ApJ...692L..67A} {692, L67}

\bibitem[\protect\citeauthoryear{{Audard} et~al.,}{{Audard}
  et~al.}{2014}]{audard2014}
{Audard} M.,  et~al., 2014, \mn@doi [Protostars and Planets VI]
  {10.2458/azu_uapress_9780816531240-ch017}, \href
  {http://adsabs.harvard.edu/abs/2014prpl.conf..387A} {pp 387--410}

\bibitem[\protect\citeauthoryear{{Bae}, {Hartmann}, {Zhu}  \& {Nelson}}{{Bae}
  et~al.}{2014}]{Bae2014}
{Bae} J.,  {Hartmann} L.,  {Zhu} Z.,   {Nelson} R.~P.,  2014, \mn@doi [\apj]
  {10.1088/0004-637X/795/1/61}, \href
  {http://adsabs.harvard.edu/abs/2014ApJ...795...61B} {795, 61}

\bibitem[\protect\citeauthoryear{{Bal\'azs}, {Eisl\"offel}, {Holl}, {Kelemen}
  \& {Kun}}{{Bal\'azs} et~al.}{1992}]{balazs1992}
{Bal\'azs} L.~G.,  {Eisl\"offel} J.,  {Holl} A.,  {Kelemen} J.,   {Kun} M.,
  1992, \aap, \href {http://adsabs.harvard.edu/abs/1992A%26A...255..281B} {255,
  281}

\bibitem[\protect\citeauthoryear{{Bal{\'a}zs}, {{\'A}brah{\'a}m}, {Kun},
  {Kelemen}  \& {T{\'o}th}}{{Bal{\'a}zs} et~al.}{2004}]{balazs2004}
{Bal{\'a}zs} L.~G.,  {{\'A}brah{\'a}m} P.,  {Kun} M.,  {Kelemen} J.,
  {T{\'o}th} L.~V.,  2004, \mn@doi [\aap] {10.1051/0004-6361:20047059}, \href
  {http://adsabs.harvard.edu/abs/2004A%26A...425..133B} {425, 133}

\bibitem[\protect\citeauthoryear{{Beck}}{{Beck}}{2007}]{beck2007}
{Beck} T.~L.,  2007, \mn@doi [\aj] {10.1086/511784}, \href
  {http://adsabs.harvard.edu/abs/2007AJ....133.1673B} {133, 1673}

\bibitem[\protect\citeauthoryear{{Bell} \& {Lin}}{{Bell} \&
  {Lin}}{1994}]{Bell1994}
{Bell} K.~R.,  {Lin} D.~N.~C.,  1994, \mn@doi [\apj] {10.1086/174206}, \href
  {http://adsabs.harvard.edu/abs/1994ApJ...427..987B} {427, 987}

\bibitem[\protect\citeauthoryear{{Bell}, {Lin}, {Hartmann}  \& {Kenyon}}{{Bell}
  et~al.}{1995}]{Bell1995}
{Bell} K.~R.,  {Lin} D.~N.~C.,  {Hartmann} L.~W.,   {Kenyon} S.~J.,  1995,
  \mn@doi [\apj] {10.1086/175612}, \href
  {http://adsabs.harvard.edu/abs/1995ApJ...444..376B} {444, 376}

\bibitem[\protect\citeauthoryear{{\swap{Burgo}{del }}, {Laureijs},
  {{\'A}brah{\'a}m}  \& {Kiss}}{{\swap{Burgo}{del }} et~al.}{2003}]{Burgo2003}
{\swap{Burgo}{del }} C.,  {Laureijs} R.~J.,  {{\'A}brah{\'a}m} P.,   {Kiss} C.,
   2003, \mn@doi [\mnras] {10.1046/j.1365-2966.2003.07081.x}, \href
  {http://adsabs.harvard.edu/abs/2003MNRAS.346..403D} {346, 403}

\bibitem[\protect\citeauthoryear{{Cardelli}, {Clayton}  \& {Mathis}}{{Cardelli}
  et~al.}{1989}]{cardelli1989}
{Cardelli} J.~A.,  {Clayton} G.~C.,   {Mathis} J.~S.,  1989, \mn@doi [\apj]
  {10.1086/167900}, \href {http://adsabs.harvard.edu/abs/1989ApJ...345..245C}
  {345, 245}

\bibitem[\protect\citeauthoryear{{Cesarsky} et~al.,}{{Cesarsky}
  et~al.}{1996}]{cesarsky1996}
{Cesarsky} C.~J.,  et~al., 1996, \aap, \href
  {http://adsabs.harvard.edu/abs/1996A%26A...315L..32C} {315, L32}

\bibitem[\protect\citeauthoryear{{Chambers} et~al.,}{{Chambers}
  et~al.}{2016}]{Chambers2016}
{Chambers} K.~C.,  et~al., 2016, preprint, \href
  {http://adsabs.harvard.edu/abs/2016arXiv161205560C} {} (\mn@eprint {arXiv}
  {1612.05560})

\bibitem[\protect\citeauthoryear{{Cohen}}{{Cohen}}{1980}]{cohen1980}
{Cohen} M.,  1980, \mn@doi [\aj] {10.1086/112630}, \href
  {http://adsabs.harvard.edu/abs/1980AJ.....85...29C} {85, 29}

\bibitem[\protect\citeauthoryear{{Connelley} \& {Reipurth}}{{Connelley} \&
  {Reipurth}}{2018}]{Connelley2018}
{Connelley} M.~S.,  {Reipurth} B.,  2018, \mn@doi [\apj]
  {10.3847/1538-4357/aaba7b}, \href
  {http://adsabs.harvard.edu/abs/2018ApJ...861..145C} {861, 145}

\bibitem[\protect\citeauthoryear{{Contreras Pe{\~n}a} et~al.,}{{Contreras
  Pe{\~n}a} et~al.}{2017a}]{contreras1}
{Contreras Pe{\~n}a} C.,  et~al., 2017a, \mn@doi [\mnras]
  {10.1093/mnras/stw2801}, \href
  {http://adsabs.harvard.edu/abs/2017MNRAS.465.3011C} {465, 3011}

\bibitem[\protect\citeauthoryear{{Contreras Pe{\~n}a} et~al.,}{{Contreras
  Pe{\~n}a} et~al.}{2017b}]{contreras2}
{Contreras Pe{\~n}a} C.,  et~al., 2017b, \mn@doi [\mnras]
  {10.1093/mnras/stw2802}, \href
  {http://adsabs.harvard.edu/abs/2017MNRAS.465.3039C} {465, 3039}

\bibitem[\protect\citeauthoryear{{Cutri} et~al.,}{{Cutri} et~al.}{2003}]{2MASS}
{Cutri} R.~M.,  et~al., 2003, VizieR Online Data Catalog, \href
  {http://adsabs.harvard.edu/abs/2003yCat.2246....0C} {2246}

\bibitem[\protect\citeauthoryear{{Cutri} et~al.,}{{Cutri}
  et~al.}{2013}]{wise2013}
{Cutri} R.~M.,  et~al., 2013, Technical report, {Explanatory Supplement to the
  AllWISE Data Release Products}.
IPAC/California Institute of Technology

\bibitem[\protect\citeauthoryear{{Diolaiti}, {Bendinelli}, {Bonaccini},
  {Close}, {Currie}  \& {Parmeggiani}}{{Diolaiti} et~al.}{2000}]{Diolaiti2000}
{Diolaiti} E.,  {Bendinelli} O.,  {Bonaccini} D.,  {Close} L.,  {Currie} D.,
  {Parmeggiani} G.,  2000, \mn@doi [\aaps] {10.1051/aas:2000305}, \href
  {http://adsabs.harvard.edu/abs/2000A%26AS..147..335D} {147, 335}

\bibitem[\protect\citeauthoryear{{Dunham} et~al.,}{{Dunham}
  et~al.}{2013}]{Dunham2013}
{Dunham} M.~M.,  et~al., 2013, \mn@doi [\aj] {10.1088/0004-6256/145/4/94},
  \href {http://adsabs.harvard.edu/abs/2013AJ....145...94D} {145, 94}

\bibitem[\protect\citeauthoryear{{Engelbracht} et~al.,}{{Engelbracht}
  et~al.}{2007}]{Engelbracht2007}
{Engelbracht} C.~W.,  et~al., 2007, \mn@doi [\pasp] {10.1086/521881}, \href
  {http://adsabs.harvard.edu/abs/2007PASP..119..994E} {119, 994}

\bibitem[\protect\citeauthoryear{{Evans} II et~al.,}{{Evans}
  et~al.}{2003}]{Evans2003}
{Evans} II N.~J.,  et~al., 2003, \mn@doi [\pasp] {10.1086/376697}, \href
  {http://adsabs.harvard.edu/abs/2003PASP..115..965E} {115, 965}

\bibitem[\protect\citeauthoryear{{Fazio} et~al.,}{{Fazio}
  et~al.}{2004}]{fazio2004}
{Fazio} G.~G.,  et~al., 2004, \mn@doi [\apjs] {10.1086/422843}, \href
  {http://adsabs.harvard.edu/abs/2004ApJS..154...10F} {154, 10}

\bibitem[\protect\citeauthoryear{{Fischer} et~al.,}{{Fischer}
  et~al.}{2012}]{fischer2012}
{Fischer} W.~J.,  et~al., 2012, \mn@doi [\apj] {10.1088/0004-637X/756/1/99},
  \href {http://adsabs.harvard.edu/abs/2012ApJ...756...99F} {756, 99}

\bibitem[\protect\citeauthoryear{{Furlan} et~al.,}{{Furlan}
  et~al.}{2016}]{furlan2016}
{Furlan} E.,  et~al., 2016, \mn@doi [\apjs] {10.3847/0067-0049/224/1/5}, \href
  {http://adsabs.harvard.edu/abs/2016ApJS..224....5F} {224, 5}

\bibitem[\protect\citeauthoryear{{Gabriel}, {Acosta-Pulido}, {Heinrichsen},
  {Morris}  \& {Tai}}{{Gabriel} et~al.}{1997}]{Gabriel1997}
{Gabriel} C.,  {Acosta-Pulido} J.,  {Heinrichsen} I.,  {Morris} H.,   {Tai}
  W.-M.,  1997, in {Hunt} G.,  {Payne} H.,  eds,  Astronomical Society of the
  Pacific Conference Series Vol. 125, Astronomical Data Analysis Software and
  Systems VI. p.~108

\bibitem[\protect\citeauthoryear{{Gaia Collaboration} et~al.,}{{Gaia
  Collaboration} et~al.}{2018}]{gaia2018}
{Gaia Collaboration} et~al., 2018, \mn@doi [\aap]
  {10.1051/0004-6361/201833051}, \href
  {http://adsabs.harvard.edu/abs/2018A%26A...616A...1G} {616, A1}

\bibitem[\protect\citeauthoryear{{Gordon} et~al.,}{{Gordon}
  et~al.}{2007}]{Gordon2007}
{Gordon} K.~D.,  et~al., 2007, \mn@doi [\pasp] {10.1086/522675}, \href
  {http://adsabs.harvard.edu/abs/2007PASP..119.1019G} {119, 1019}

\bibitem[\protect\citeauthoryear{{Greene}, {Wilking}, {Andre}, {Young}  \&
  {Lada}}{{Greene} et~al.}{1994}]{greene94}
{Greene} T.~P.,  {Wilking} B.~A.,  {Andre} P.,  {Young} E.~T.,   {Lada} C.~J.,
  1994, \mn@doi [\apj] {10.1086/174763}, \href
  {http://adsabs.harvard.edu/abs/1994ApJ...434..614G} {434, 614}

\bibitem[\protect\citeauthoryear{{Hartmann}}{{Hartmann}}{1998}]{Hartmann1998}
{Hartmann} L.,  1998, {Accretion Processes in Star Formation}.
Cambridge astrophysics series, Cambridge University Press

\bibitem[\protect\citeauthoryear{{Johnstone}, {Hendricks}, {Herczeg}  \&
  {Bruderer}}{{Johnstone} et~al.}{2013}]{Johnstone2013}
{Johnstone} D.,  {Hendricks} B.,  {Herczeg} G.~J.,   {Bruderer} S.,  2013,
  \mn@doi [\apj] {10.1088/0004-637X/765/2/133}, \href
  {http://adsabs.harvard.edu/abs/2013ApJ...765..133J} {765, 133}

\bibitem[\protect\citeauthoryear{{Kawada} et~al.,}{{Kawada}
  et~al.}{2007}]{kawada2007}
{Kawada} M.,  et~al., 2007, \mn@doi [\pasj] {10.1093/pasj/59.sp2.S389}, \href
  {http://adsabs.harvard.edu/abs/2007PASJ...59S.389K} {59, S389}

\bibitem[\protect\citeauthoryear{{Kim} et~al.,}{{Kim} et~al.}{2015}]{kim2015}
{Kim} J.,  et~al., 2015, \mn@doi [\apjs] {10.1088/0067-0049/218/1/5}, \href
  {http://adsabs.harvard.edu/abs/2015ApJS..218....5K} {218, 5}

\bibitem[\protect\citeauthoryear{{K{\'o}sp{\'a}l}, {{\'A}brah{\'a}m}, {Prusti},
  {Acosta-Pulido}, {Hony}, {Mo{\'o}r}  \& {Siebenmorgen}}{{K{\'o}sp{\'a}l}
  et~al.}{2007}]{kospal2007}
{K{\'o}sp{\'a}l} {\'A}.,  {{\'A}brah{\'a}m} P.,  {Prusti} T.,  {Acosta-Pulido}
  J.,  {Hony} S.,  {Mo{\'o}r} A.,   {Siebenmorgen} R.,  2007, \mn@doi [\aap]
  {10.1051/0004-6361:20066108}, \href
  {http://adsabs.harvard.edu/abs/2007A%26A...470..211K} {470, 211}

\bibitem[\protect\citeauthoryear{{K{\'o}sp{\'a}l} et~al.,}{{K{\'o}sp{\'a}l}
  et~al.}{2016}]{kospal2016}
{K{\'o}sp{\'a}l} {\'A}.,  et~al., 2016, \mn@doi [\aap]
  {10.1051/0004-6361/201528061}, \href
  {http://adsabs.harvard.edu/abs/2016A%26A...596A..52K} {596, A52}

\bibitem[\protect\citeauthoryear{{K{\'o}sp{\'a}l}, {{\'A}brah{\'a}m},
  {Westhues}  \& {Haas}}{{K{\'o}sp{\'a}l} et~al.}{2017a}]{kospal2017b}
{K{\'o}sp{\'a}l} {\'A}.,  {{\'A}brah{\'a}m} P.,  {Westhues} C.,   {Haas} M.,
  2017a, \mn@doi [\aap] {10.1051/0004-6361/201629447}, \href
  {http://adsabs.harvard.edu/abs/2017A%26A...597L..10K} {597, L10}

\bibitem[\protect\citeauthoryear{{K{\'o}sp{\'a}l} et~al.,}{{K{\'o}sp{\'a}l}
  et~al.}{2017b}]{kospal2017a}
{K{\'o}sp{\'a}l} {\'A}.,  et~al., 2017b, \mn@doi [\apj]
  {10.3847/1538-4357/aa7683}, \href
  {http://adsabs.harvard.edu/abs/2017ApJ...843...45K} {843, 45}

\bibitem[\protect\citeauthoryear{{Kristensen} \& {Dunham}}{{Kristensen} \&
  {Dunham}}{2018}]{Kristensen2018}
{Kristensen} L.~E.,  {Dunham} M.~M.,  2018, preprint, \href
  {http://adsabs.harvard.edu/abs/2018arXiv180711262K} {} (\mn@eprint {arXiv}
  {1807.11262})

\bibitem[\protect\citeauthoryear{{Kuffmeier}, {Frimann}, {Jensen}  \&
  {Haugb{\o}lle}}{{Kuffmeier} et~al.}{2018}]{Kuffmeier2018}
{Kuffmeier} M.,  {Frimann} S.,  {Jensen} S.~S.,   {Haugb{\o}lle} T.,  2018,
  \mn@doi [\mnras] {10.1093/mnras/sty024}, \href
  {http://adsabs.harvard.edu/abs/2018MNRAS.475.2642K} {475, 2642}

\bibitem[\protect\citeauthoryear{{Kun} \& {Prusti}}{{Kun} \&
  {Prusti}}{1993}]{KP93}
{Kun} M.,  {Prusti} T.,  1993, \aap, \href
  {http://adsabs.harvard.edu/abs/1993A%26A...272..235K} {272, 235}

\bibitem[\protect\citeauthoryear{{Kun}, {Balog}, {Kenyon}, {Mamajek}  \&
  {Gutermuth}}{{Kun} et~al.}{2009}]{kun2009}
{Kun} M.,  {Balog} Z.,  {Kenyon} S.~J.,  {Mamajek} E.~E.,   {Gutermuth} R.~A.,
  2009, \mn@doi [\apjs] {10.1088/0067-0049/185/2/451}, \href
  {http://adsabs.harvard.edu/abs/2009ApJS..185..451K} {185, 451}

\bibitem[\protect\citeauthoryear{{Laureijs}, {Klaas}, {Richards}, {Schulz}  \&
  {Abraham}}{{Laureijs} et~al.}{2003}]{Laureijs2003}
{Laureijs} R.~J.,  {Klaas} U.,  {Richards} P.~J.,  {Schulz} B.,   {Abraham} P.,
   2003, {The ISO Handbook, Volume IV - PHT - The Imaging Photo-Polarimeter}.
European Space Agency

\bibitem[\protect\citeauthoryear{{Mainzer} et~al.,}{{Mainzer}
  et~al.}{2014}]{Mainzer2014}
{Mainzer} A.,  et~al., 2014, \mn@doi [\apj] {10.1088/0004-637X/792/1/30}, \href
  {http://adsabs.harvard.edu/abs/2014ApJ...792...30M} {792, 30}

\bibitem[\protect\citeauthoryear{{Makovoz} \& {Marleau}}{{Makovoz} \&
  {Marleau}}{2005}]{Makovoz2005}
{Makovoz} D.,  {Marleau} F.~R.,  2005, \mn@doi [\pasp] {10.1086/432977}, \href
  {http://adsabs.harvard.edu/abs/2005PASP..117.1113M} {117, 1113}

\bibitem[\protect\citeauthoryear{{Marton} et~al.,}{{Marton}
  et~al.}{2017}]{Marton2017}
{Marton} G.,  et~al., 2017, preprint, \href
  {http://adsabs.harvard.edu/abs/2017arXiv170505693M} {} (\mn@eprint {arXiv}
  {1705.05693})

\bibitem[\protect\citeauthoryear{{Meyer}, {Calvet}  \& {Hillenbrand}}{{Meyer}
  et~al.}{1997}]{Meyer}
{Meyer} M.~R.,  {Calvet} N.,   {Hillenbrand} L.~A.,  1997, \mn@doi [\aj]
  {10.1086/118474}, \href {http://adsabs.harvard.edu/abs/1997AJ....114..288M}
  {114, 288}

\bibitem[\protect\citeauthoryear{{Murakami} et~al.,}{{Murakami}
  et~al.}{2007}]{akari}
{Murakami} H.,  et~al., 2007, \mn@doi [\pasj] {10.1093/pasj/59.sp2.S369}, \href
  {http://adsabs.harvard.edu/abs/2007PASJ...59S.369M} {59, S369}

\bibitem[\protect\citeauthoryear{{Myers} \& {Ladd}}{{Myers} \&
  {Ladd}}{1993}]{Myers1993}
{Myers} P.~C.,  {Ladd} E.~F.,  1993, \mn@doi [\apjl] {10.1086/186956}, \href
  {http://adsabs.harvard.edu/abs/1993ApJ...413L..47M} {413, L47}

\bibitem[\protect\citeauthoryear{{Nikoli{\'c}}, {Johansson}  \&
  {Harju}}{{Nikoli{\'c}} et~al.}{2003}]{nikolic2003}
{Nikoli{\'c}} S.,  {Johansson} L.~E.~B.,   {Harju} J.,  2003, \mn@doi [\aap]
  {10.1051/0004-6361:20031102}, \href
  {http://adsabs.harvard.edu/abs/2003A%26A...409..941N} {409, 941}

\bibitem[\protect\citeauthoryear{{Ninan} et~al.,}{{Ninan}
  et~al.}{2015}]{Ninan2015}
{Ninan} J.~P.,  et~al., 2015, \mn@doi [\apj] {10.1088/0004-637X/815/1/4}, \href
  {http://adsabs.harvard.edu/abs/2015ApJ...815....4N} {815, 4}

\bibitem[\protect\citeauthoryear{{Onaka} et~al.,}{{Onaka}
  et~al.}{2007}]{onaka2007}
{Onaka} T.,  et~al., 2007, \mn@doi [\pasj] {10.1093/pasj/59.sp2.S401}, \href
  {http://adsabs.harvard.edu/abs/2007PASJ...59S.401O} {59, S401}

\bibitem[\protect\citeauthoryear{{Onozato}, {Ita}, {Ono}, {Fukagawa},
  {Yanagisawa}, {Izumiura}, {Nakada}  \& {Matsunaga}}{{Onozato}
  et~al.}{2015}]{onozato2015}
{Onozato} H.,  {Ita} Y.,  {Ono} K.,  {Fukagawa} M.,  {Yanagisawa} K.,
  {Izumiura} H.,  {Nakada} Y.,   {Matsunaga} N.,  2015, \mn@doi [\pasj]
  {10.1093/pasj/psv008}, \href
  {http://adsabs.harvard.edu/abs/2015PASJ...67...39O} {67, 39}

\bibitem[\protect\citeauthoryear{{Ott} et~al.,}{{Ott} et~al.}{1997}]{Ott1997}
{Ott} S.,  et~al., 1997, in {Hunt} G.,  {Payne} H.,  eds,  Astronomical Society
  of the Pacific Conference Series Vol. 125, Astronomical Data Analysis
  Software and Systems VI. p.~34

\bibitem[\protect\citeauthoryear{{Paladini} \& {Noriega-Crespo}}{{Paladini} \&
  {Noriega-Crespo}}{2009}]{paladini2009}
{Paladini} R.,  {Noriega-Crespo} A.,  2009, Characterization of MIPS 70\,$\mu$m
  flux non-linearity,
  \url{https://irsa.ipac.caltech.edu/data/SPITZER/docs/files/spitzer/Non_linearity_70um_v2.pdf}

\bibitem[\protect\citeauthoryear{{Pattle} et~al.,}{{Pattle}
  et~al.}{2017}]{Pattle2017}
{Pattle} K.,  et~al., 2017, \mn@doi [\mnras] {10.1093/mnras/stw2648}, \href
  {http://adsabs.harvard.edu/abs/2017MNRAS.464.4255P} {464, 4255}

\bibitem[\protect\citeauthoryear{{Reipurth}, {Rodr{\'{\i}}guez}, {Anglada}  \&
  {Bally}}{{Reipurth} et~al.}{2004}]{reipurth2004}
{Reipurth} B.,  {Rodr{\'{\i}}guez} L.~F.,  {Anglada} G.,   {Bally} J.,  2004,
  \mn@doi [\aj] {10.1086/381062}, \href
  {http://adsabs.harvard.edu/abs/2004AJ....127.1736R} {127, 1736}

\bibitem[\protect\citeauthoryear{{Rieke} et~al.,}{{Rieke}
  et~al.}{2004}]{Rieke2004}
{Rieke} G.~H.,  et~al., 2004, \mn@doi [\apjs] {10.1086/422717}, \href
  {http://adsabs.harvard.edu/abs/2004ApJS..154...25R} {154, 25}

\bibitem[\protect\citeauthoryear{{Rosvick} \& {Davidge}}{{Rosvick} \&
  {Davidge}}{1995}]{RD95}
{Rosvick} J.~M.,  {Davidge} T.~J.,  1995, \mn@doi [\pasp] {10.1086/133514},
  \href {http://adsabs.harvard.edu/abs/1995PASP..107...49R} {107, 49}

\bibitem[\protect\citeauthoryear{{Safron} et~al.,}{{Safron}
  et~al.}{2015}]{safron2015}
{Safron} E.~J.,  et~al., 2015, \mn@doi [\apjl] {10.1088/2041-8205/800/1/L5},
  \href {http://adsabs.harvard.edu/abs/2015ApJ...800L...5S} {800, L5}

\bibitem[\protect\citeauthoryear{{Sato} \& {Fukui}}{{Sato} \&
  {Fukui}}{1989}]{sato1989}
{Sato} F.,  {Fukui} Y.,  1989, \mn@doi [\apj] {10.1086/167748}, \href
  {http://adsabs.harvard.edu/abs/1989ApJ...343..773S} {343, 773}

\bibitem[\protect\citeauthoryear{{Siess}, {Dufour}  \& {Forestini}}{{Siess}
  et~al.}{2000}]{siess2000}
{Siess} L.,  {Dufour} E.,   {Forestini} M.,  2000, \aap, \href
  {http://adsabs.harvard.edu/abs/2000A%26A...358..593S} {358, 593}

\bibitem[\protect\citeauthoryear{{Suresh}, {Dunham}, {Arce}, {Evans}, {Bourke},
  {Merello}  \& {Wu}}{{Suresh} et~al.}{2016}]{suresh2016}
{Suresh} A.,  {Dunham} M.~M.,  {Arce} H.~G.,  {Evans} II N.~J.,  {Bourke}
  T.~L.,  {Merello} M.,   {Wu} J.,  2016, \mn@doi [\aj]
  {10.3847/0004-6256/152/2/36}, \href
  {http://cdsads.u-strasbg.fr/abs/2016AJ....152...36S} {152, 36}

\bibitem[\protect\citeauthoryear{{Vorobyov} \& {Basu}}{{Vorobyov} \&
  {Basu}}{2006}]{Vorobyov2006}
{Vorobyov} E.~I.,  {Basu} S.,  2006, \mn@doi [\apj] {10.1086/507320}, \href
  {http://adsabs.harvard.edu/abs/2006ApJ...650..956V} {650, 956}

\bibitem[\protect\citeauthoryear{{Vorobyov} \& {Basu}}{{Vorobyov} \&
  {Basu}}{2015}]{Vorobyov2015}
{Vorobyov} E.~I.,  {Basu} S.,  2015, \mn@doi [\apj]
  {10.1088/0004-637X/805/2/115}, \href
  {http://adsabs.harvard.edu/abs/2015ApJ...805..115V} {805, 115}

\bibitem[\protect\citeauthoryear{{White} \& {Hillenbrand}}{{White} \&
  {Hillenbrand}}{2004}]{white2004}
{White} R.~J.,  {Hillenbrand} L.~A.,  2004, \mn@doi [\apj] {10.1086/425115},
  \href {http://adsabs.harvard.edu/abs/2004ApJ...616..998W} {616, 998}

\bibitem[\protect\citeauthoryear{{Whitney}, {Wood}, {Bjorkman}  \&
  {Wolff}}{{Whitney} et~al.}{2003}]{whitney2003}
{Whitney} B.~A.,  {Wood} K.,  {Bjorkman} J.~E.,   {Wolff} M.~J.,  2003, \mn@doi
  [\apj] {10.1086/375415}, \href
  {http://adsabs.harvard.edu/abs/2003ApJ...591.1049W} {591, 1049}

\bibitem[\protect\citeauthoryear{{Wright} et~al.,}{{Wright}
  et~al.}{2010}]{Wright2010}
{Wright} E.~L.,  et~al., 2010, \mn@doi [\aj] {10.1088/0004-6256/140/6/1868},
  \href {http://adsabs.harvard.edu/abs/2010AJ....140.1868W} {140, 1868}

\bibitem[\protect\citeauthoryear{{Xue}, {Jiang}, {Gao}, {Liu}, {Wang}  \&
  {Li}}{{Xue} et~al.}{2016}]{xue2016}
{Xue} M.,  {Jiang} B.~W.,  {Gao} J.,  {Liu} J.,  {Wang} S.,   {Li} A.,  2016,
  \mn@doi [\apjs] {10.3847/0067-0049/224/2/23}, \href
  {http://adsabs.harvard.edu/abs/2016ApJS..224...23X} {224, 23}

\bibitem[\protect\citeauthoryear{{Yamamura} \& {the \textit{Akari\/}
  team}}{{Yamamura} \& {the \textit{Akari\/} team}}{2016}]{yamamura2016}
{Yamamura} I.,  {the \textit{Akari\/} team} 2016, AKARI-FIS Bright Source
  Catalogue Public Version 2.,
  \url{http://www.ir.isas.jaxa.jp/AKARI/Archive/Catalogues/FISBSCv2/}

\bibitem[\protect\citeauthoryear{{Yamashita}, {Murata}, {Egusa}, {Usui}  \&
  {Yamamura}}{{Yamashita} et~al.}{2016}]{yamashita2016}
{Yamashita} T.,  {Murata} K.,  {Egusa} F.,  {Usui} F.,   {Yamamura} I.,  2016,
  AKARI/IRC Pointed Observation Images (Post-Helium Mission),
  \url{http://www.ir.isas.jaxa.jp/AKARI/Archive/Images/IRC_Images_P3/}

\bibitem[\protect\citeauthoryear{{Zhu}, {Hartmann}  \& {Gammie}}{{Zhu}
  et~al.}{2009}]{Zhu2009}
{Zhu} Z.,  {Hartmann} L.,   {Gammie} C.,  2009, \mn@doi [\apj]
  {10.1088/0004-637X/694/2/1045}, \href
  {http://adsabs.harvard.edu/abs/2009ApJ...694.1045Z} {694, 1045}

\makeatother
\end{thebibliography}

\bsp	
\label{lastpage}

\end{document}